\documentclass[12pt,twoside,a4paper]{book}

\usepackage[a4paper,
            inner=3cm,
            outer=2.5cm,
            top=3cm,
            bottom=3cm,
            bindingoffset=1cm]{geometry}

\def\TheTitle{Small-Angle Neutron Scattering \\--\\ SANS}                        
\def\TheInstitute{\Large{European Spallation Source ERIC}\\ \small{Lund, Sweden}\\ \Large{J\"ulich Centre for Neutron Science 1, Forschungszentrum J\"ulich GmbH}\\ \small{J\"ulich, Germany}\\ \Large{Department of Physics and Astronomy, Uppsala University}\\ \small{Uppsala, Sweden}}       

\usepackage{pifont}
\usepackage{latexsym}
\usepackage{amsfonts}
\usepackage{amssymb}
\usepackage{amsmath}
\usepackage{amstext}

\usepackage{graphicx}
\usepackage{bm}
\usepackage{times} 

\def\vec#1{{\mathbf{#1}}}

\def\d{{\rm d}}

\def\<{\langle}
\def\>{\rangle}

\def\deg{$^{\circ}$}

\usepackage{fancyhdr}

\begin{document}

\renewcommand{\thesection}{\arabic{section}}
\renewcommand{\thesubsection}{\thesection.\arabic{subsection}}
\renewcommand{\thesubsubsection}{\thesubsection.\arabic{subsubsection}}

\renewcommand{\thefigure}{\arabic{figure}}
\renewcommand{\thetable}{\arabic{table}}
\renewcommand{\theequation}{\arabic{equation}}

\setcounter{section}{0}
\setcounter{subsection}{0}
\setcounter{figure}{0}
\setcounter{table}{0}
\setcounter{equation}{0}

\pagestyle{headings}

\begin{titlepage}
    \centering

    \vspace*{3cm}

    {\Huge\bfseries \TheTitle \par}

    \vspace{2cm}

    {\Large Sebastian Jaksch\par}

    \vspace{1cm}

    {\large \TheInstitute\par}
    \vspace{2 cm}
    {\large sebastian.jaksch@ess.eu}

    \vfill


    \vspace{1cm}

    {\large \today\par}

\end{titlepage}

\tableofcontents
\newpage
\cleardoublepage
\markboth{Small-Angle Neutron Scattering}{}


\section{Introduction}
Small-Angle Scattering (SAS) investigates structures in samples that generally range from approximately 0.5\,nm to a few 100\,nm. This can both be done for isotropic samples such as blends and liquids, as well as anisotropic samples such as quasi-crystals. In order to obtain data about that size regime scattered intensity, mostly of x-rays or neutrons, is investigated at angles from close to zero, still in the region of the primary beam up to 10\deg , depending on the wavelength of the incoming radiation.

The two primary sources for SAS experiments are x-ray (small-angle x-ray scattering, \emph{SAXS}) sources and neutron (small-angle neutron scattering, \emph{SANS}) sources, which shall be the two cases discussed here. Also scattering with electrons or other particle waves is possible, but not the main use case for the purpose of this manuscript.

For most small-angle scattering instruments, both SAXS and SANS, the science case covers the investigation of self-assembled polymeric and biological systems, multi-scale systems with large size distribution of the contained particles, solutions of (nano-)particles and soft-matter systems, protein solutions, and material science investigations. In the case of SANS this is augmented by the possibility to also investigate the spin state of the sample and hence perform investigations of the magnetic structure of the sample.

In the following sections the general setup of both SAXS and SANS instruments shall be discussed, as well as data acquisition and evaluation and preparation of the sample and the experiment in general. The information contained herein should provide sufficient information for planning and performing a SAS experiment and evaluate the gathered data.

\subsection{General concept}
\begin{figure}
\centering
\begin{minipage}[l]{0.75\textwidth}
\Large{a)}

\includegraphics[width=\textwidth]{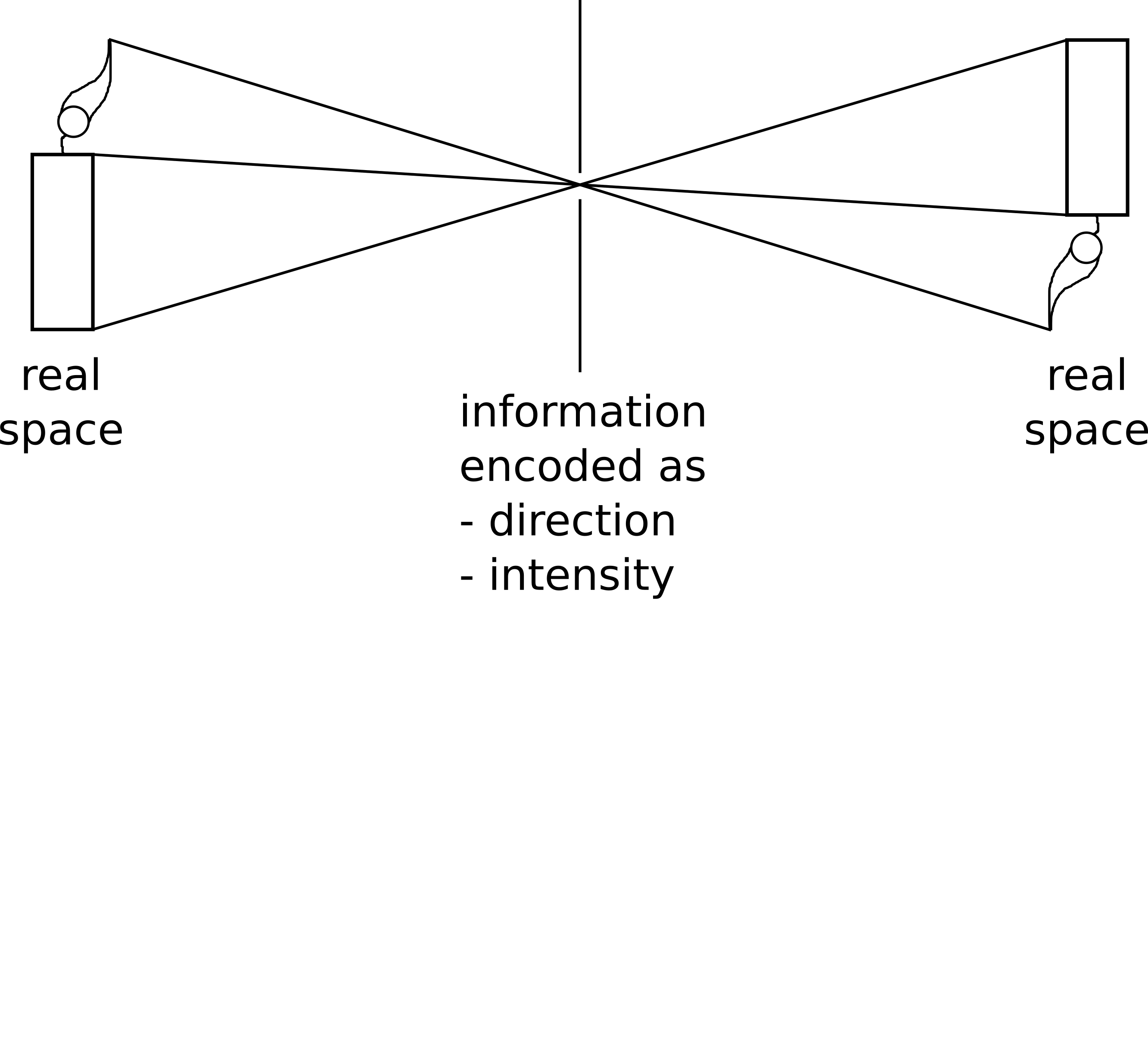}
\end{minipage}

\begin{minipage}[l]{0.75\textwidth}
\Large{b)}

\includegraphics[width=\textwidth]{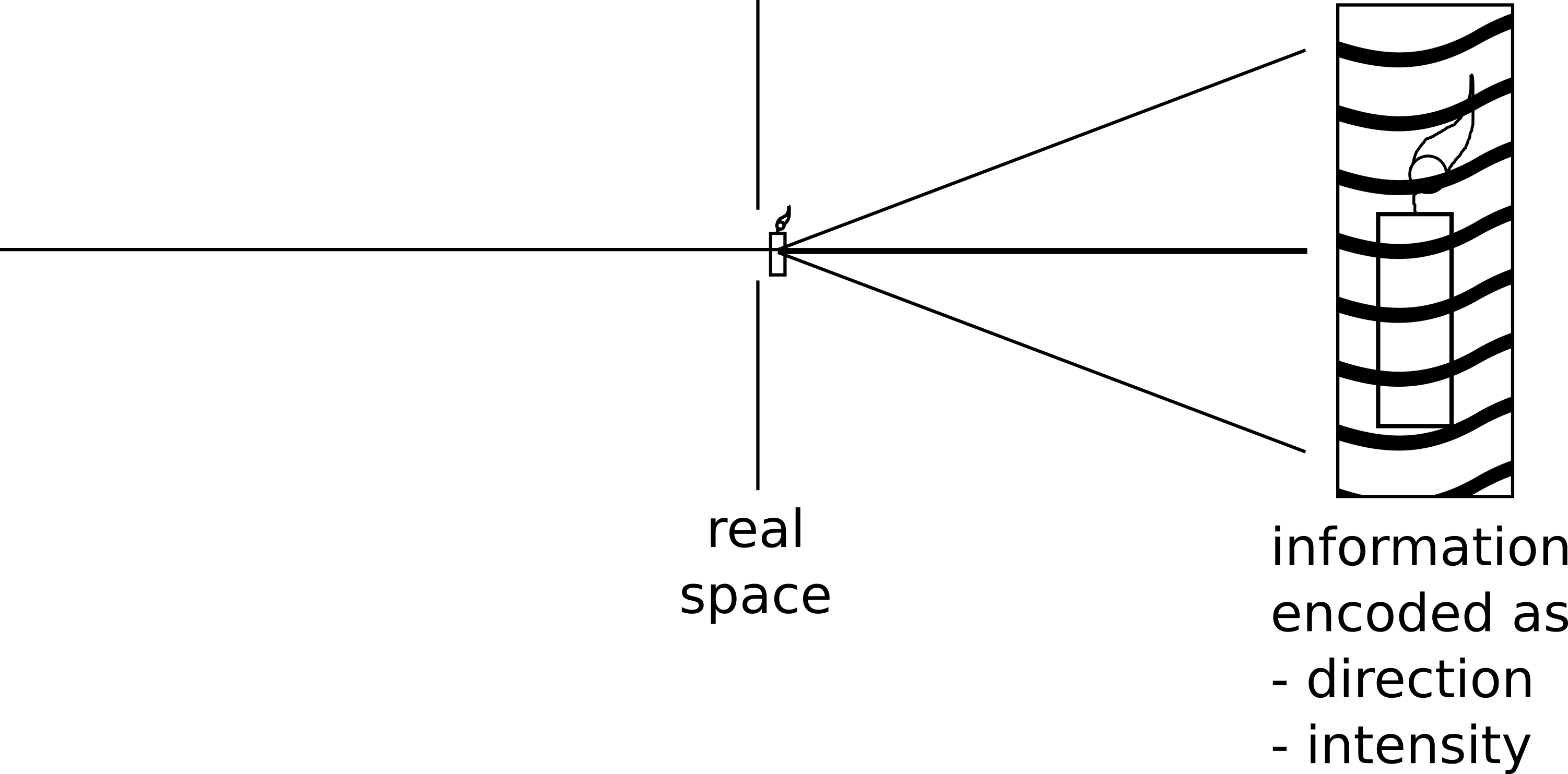}
\end{minipage}
\caption{a) Sketch of a pinhole camera and  b) a simplified SAS instrument. The encoding of the real space information is in one case done inside the pinhole, in the other case the direction (and wavelength) encoded information is directly displayed on the screen (shaded area with waves). Positioning of the screen farther away improves the angular resolution and therefore the encoded information.}
\label{fig:pinhole}
\end{figure}

All SAS experiments, irrespective of the setup used in any specific case, rely on the concept of pinhole cameras to work. Fig.\ref{fig:pinhole} illustrates the geometric concept of the interplay between pinhole cameras and SAS.

In the case of direct image projection, as done in pinhole cameras, every point of the sample (object) is mapped to a discrete point on the screen (film or detector) (see Fig. \ref{fig:pinhole}a)). The smaller the hole, the better the point-to-point mapping works, since in the ideal case only a single path between object and image is available. However, this of course comes with a penalty in intensity, since the smaller hole lets less light pass through. Due to the geometry, an image taken with a pinhole camera is always upside down. While the mathematical implications shall be discussed later on in this manuscript at this point we only want to grasp the underlying concept. The information about the object is at the beginning stored in real space. Colors (wavelength) and locations are given as points on the surface of the object. When all beams have converged to the single point that is ideally the pinhole, the information is then encoded in direction of the path (or light-beam) and the wavelength of the light. This is the change between direct and indirect space, locations and directions. When the light falls onto the screen the information is reversed again, to location and color of a spot on the screen, into direct space.

This concept is exploited by SAS (see Fig. \ref{fig:pinhole}b)). In this case only angular resolution matters, which is easy to increase simply by increasing the distance between pinhole and screen. Here, instead of using the information that has been transferred to real space again, this time the object in real space is put close to the window. This way, the information about the location of atoms and molecules in the sample is encoded into direction or indirect space. Since there should be no information about the light before the pinhole, the light needs to be collimated down to a small, point-like source with no angular divergence.

Pinhole cameras such as in Fig. \ref{fig:pinhole}a) were in use before lenses became wide spread in optical photography. One can easily imagine that a smaller aperture leads to a higher resolution at the cost of intensity. Even with optical lenses cameras operate roughly the same way. You try to offset the smaller aperture and sharper image by using a lens that focuses more light through the aperture.

For SAS, you essentially delete all information from the incoming radiation by collimating the beam (every particle/ray is going the same direction) and separating the incoming wavelength (so everything has the same energy, or in the case of visible light, color). Then the information is picked up at the sample and transferred to a screen.

\section{SAXS instruments}
\index{SAXS}
In general there are two classes of SAXS instruments. One is the laboratory type setup that can be set-up in a single laboratory with a conventional x-ray tube, or more general any metal anode setup, while the other one is a large-scale facility setup at a synchrotron that can provide higher intensities. Since the setup of both instruments differs, and also the use case is not fully identical, we shall discuss both setups separately. One thing that should be kept in mind is that the fundamental principle is identical, i.e. any experiment that can be performed at a synchrotron can also in principle be performed at a laboratory SAXS setup and is only limited in intensity. This is important for the preparation of beamtimes at a synchrotron, which in general should be thoroughly prepared in order to fully exploit all capabilities offered there.

\begin{figure}
\centering
\includegraphics[width=0.75\textwidth]{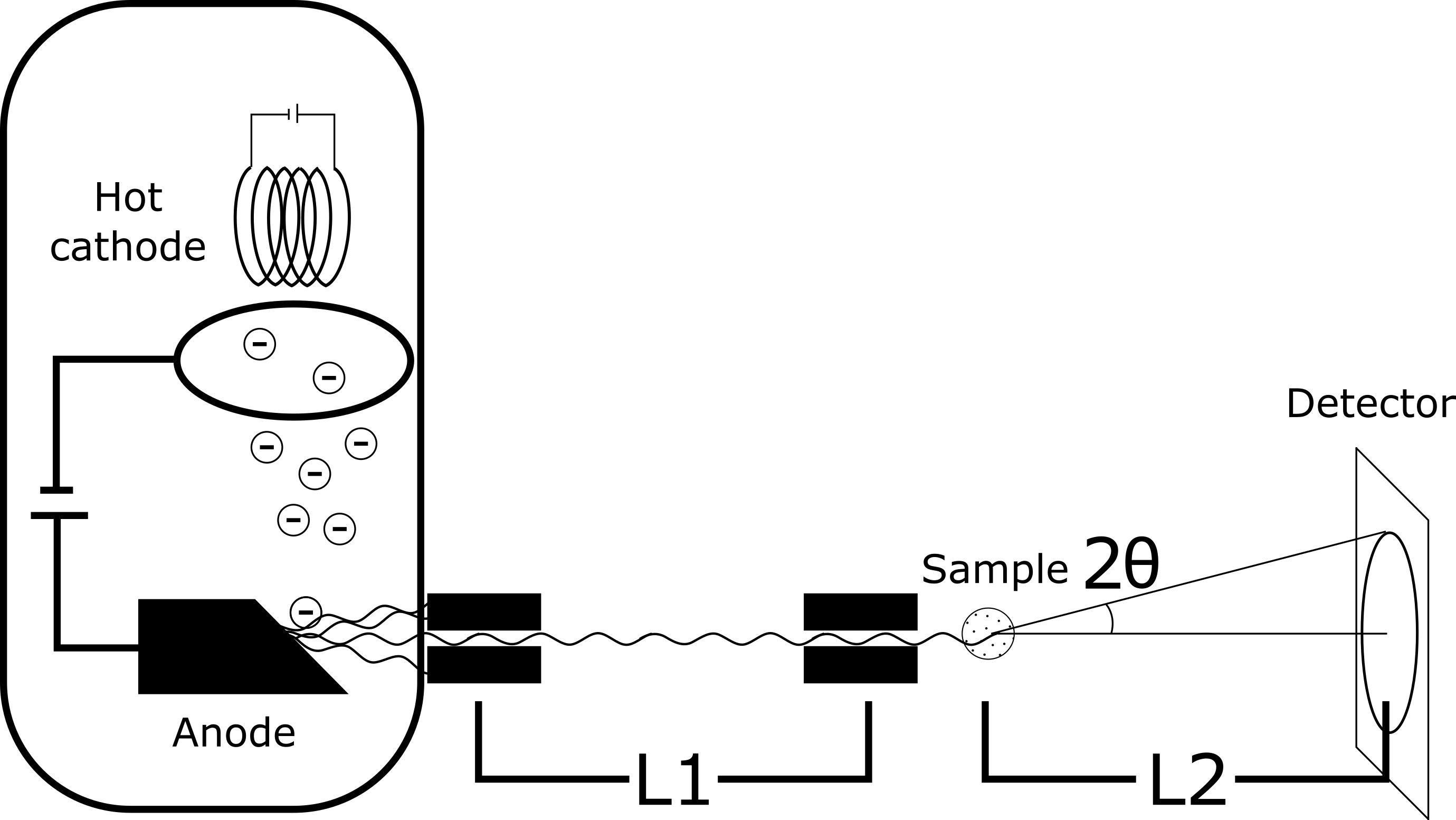}
\caption{Laboratory SAXS setup. The left box is a sketch of a x-ray tube, all the components are in vacuum. The flight path is also usually evacuated. L1 and L2 are the collimation and sample detector distance (SDD) respectively. In the case of laboratory setup those range usually from about 20 cm up to 1-2 m in modern setups. The collimation blocks for L1 and L2 are usually set up in both x and y direction to constrict the flight path, widely used openings are around 1\,mm$\times$1\,mm or below. In some setups, also a slit collimation instead of a point collimation is realized to increase the intensity.}
\label{fig:labSAXS}
\end{figure}

\subsection{Laboratory SAXS setup}
\begin{figure}
\centering
\includegraphics[width=0.75\textwidth]{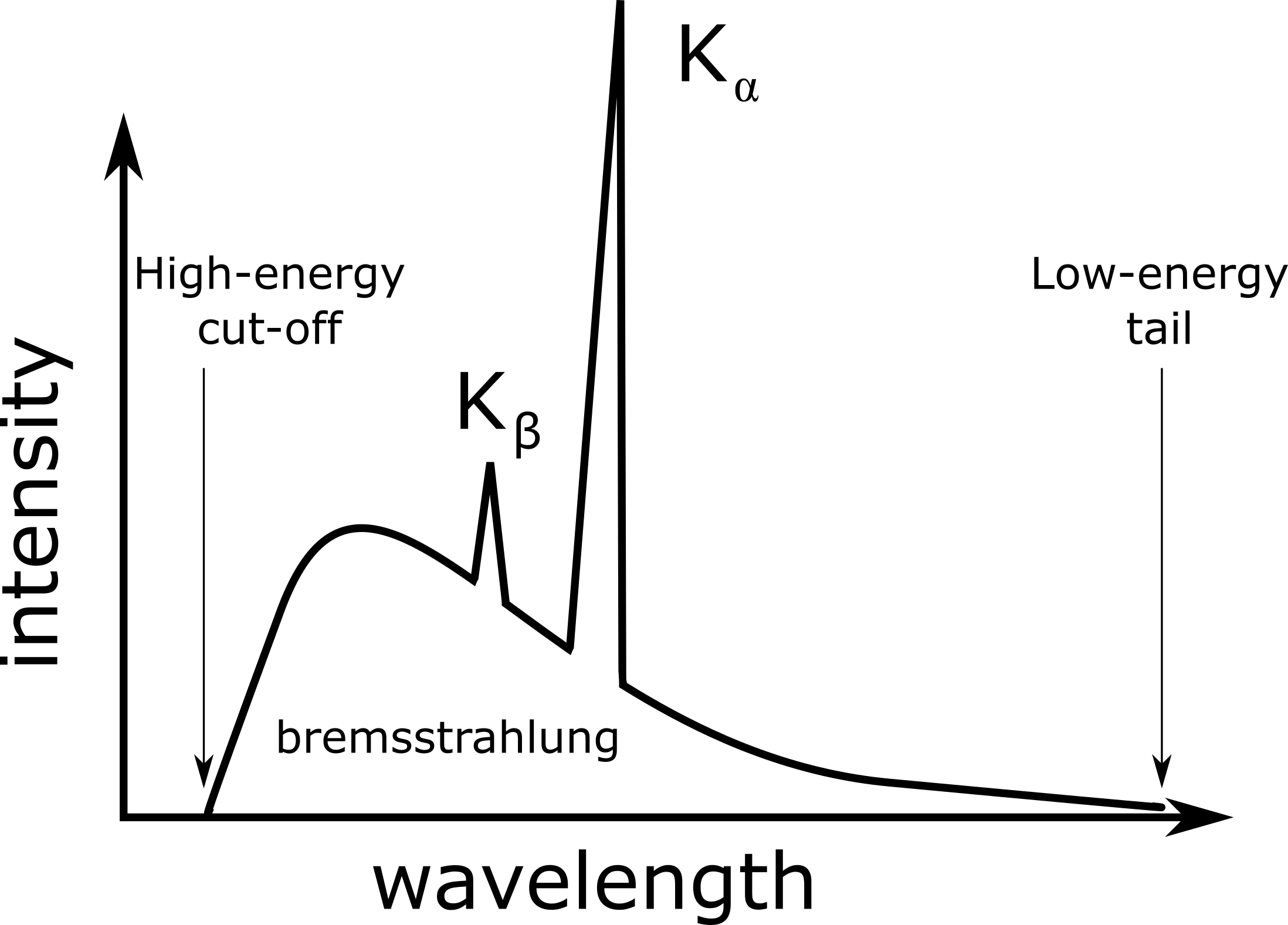}
\caption{Characteristic x-ray spectrum from a metal anode x-ray tube. The high-energy cut-off wavelength is given for the case that a single electron, fully accelerated by the voltage in the x-ray tube, deposits all its kinetic energy in a single photon. In an optimal setup this distribution is very narrow. Then the $K_\alpha$ line fully dominates the spectrum and gives a clean wavelength to perform a SAXS instrument.}
\label{fig:characteristicSpectrum}
\end{figure}

Over the years a wide range of specialized SAXS instruments has become commercially available. The oldest concepts date back to the early 20th century, right after the discovery of x-rays.\cite{guinier1964x} Most of them offer specific advantages in certain use cases, such as the measurement of isotropic samples in a Kratky Camera\cite{KratkyKamera}, or highly adaptable sample environments. Here we shall only concentrate on the basic principle of operation. A general sketch of a SAXS instrument is shown in Fig.\ref{fig:labSAXS}. The x-rays are produced in an x-ray tube and then collimated by a set of slits. Here the collimation as such is already sufficient to obtain a coherent beam, since most of the intensity of standard x-ray tubes (and essentially all metal target x-ray sources) is concentrated into the characteristic spectral lines of the target material (see Fig.\ref{fig:characteristicSpectrum}). Common materials for the target anode are copper and molybdenum, delivering wavelengths of the most intensive K-$\alpha$ lines of 1.54 \AA\, and 0.71 \AA\, respectively. Under the assumption of a usual characteristic spectrum for the anode material the x-ray tubes can be considered monochromatic sources.

In order to achieve spatial as well as wavelength coherence most x-ray tubes work with a focused beam that is as small as technically feasible. This allows very narrow collimation slits, since it is not improving the coherence, and therefore the signal-to-noise ratio, to narrow the slit further than the initial beam spot or the pixel size of the detector, whichever be smaller. This however leads to a very high energy density, why some x-ray tube designs forgo a solid anode all together and either opt for a rotating anode, where the energy of the beam spot is distributed over a larger surface or a metal-jet anode, where the material is refluxed and can therefore not heat up beyond the point of deformation and therefore also defocussing of the beam.

Some performance figures of current laboratory SAXS setups are given in Tab.\ref{tab:labSAXSparams}. It is worth noting that with the last generation of metal-jet anode setups even laboratory setups can achieve intensities comparable to what was achievable one or two decades ago at a world-class synchrotron. While this of course allows for faster measurements and smaller beam, it also means that beam damage to the sample has to be taken into account.

\begin{table}
\centering
\begin{tabular}{l|l}

Parameter & value \\ 
\hline 
\hline
SDD & 0.8-4 m \\ 
\hline 
Pixel resolution & 172$\times$172$\mu$m \\ 
\hline 
Flux & $10^7$ photons s$^{-1}$ \\ 
\hline 
wavelength $\lambda$ & 1.35 \AA \\ 
\hline 
Q-range & $4\cdot 10^{-3}$-$8\cdot 10^{-1}$ \AA$^{-1}$ \\ 
\end{tabular}

\caption{Performance parameters for state of the art laboratory SAXS setups, in this case with a liquid metal jet anode at the GALAXI instrument.\cite{kentzinger2016galaxi}}

\label{tab:labSAXSparams}
\end{table} 

\subsection{Synchrotron SAXS setups}
\begin{figure}
\centering
\includegraphics[width=0.75\textwidth]{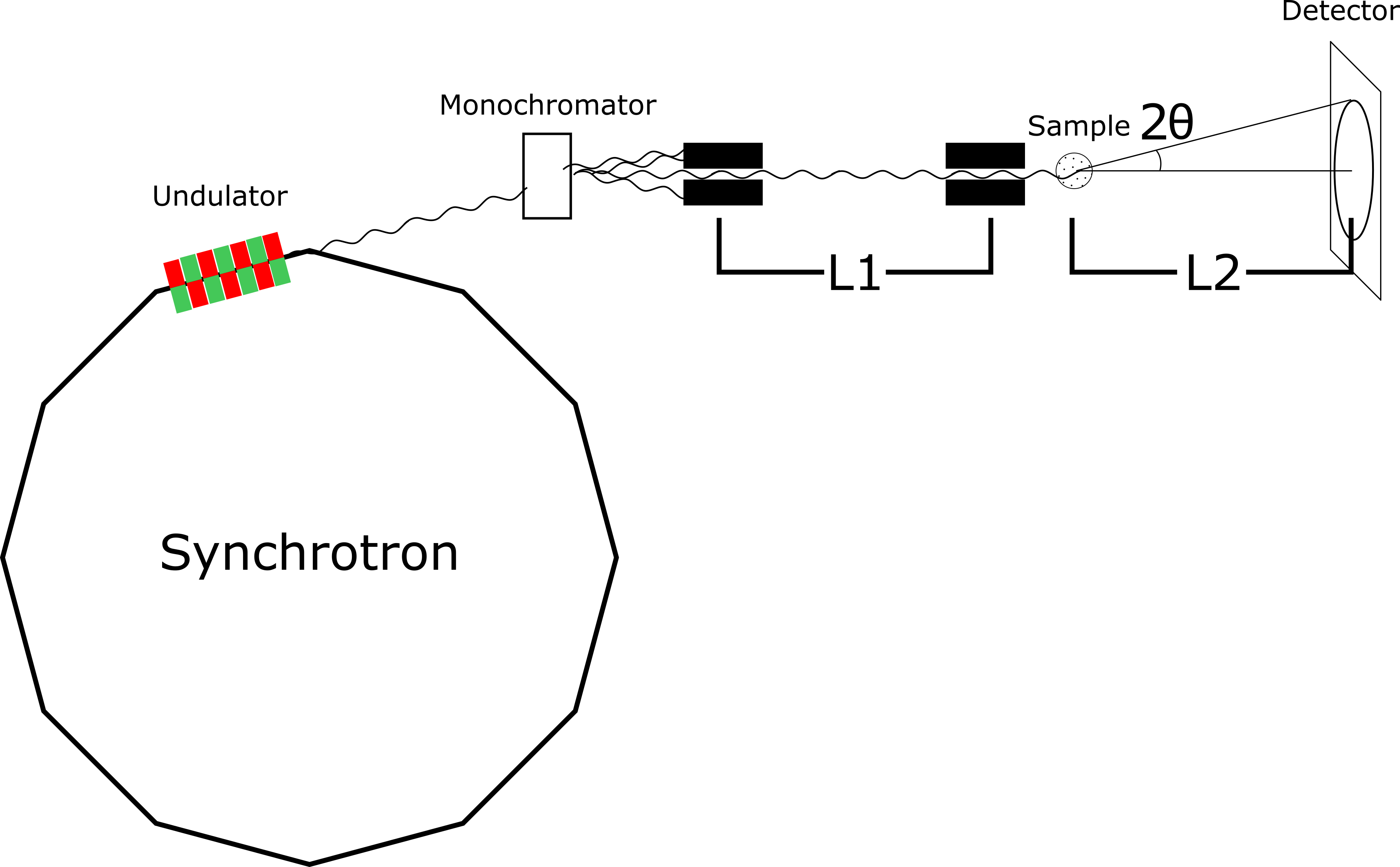}
\caption{Synchrotron SAXS setup. Here the radiation is produced in the storage ring of a synchrotron. In earlier designs, the x-rays were produced at the bending magnets in the ring (kinks in the ring here). This however lead to a wide spread of the produced wavelength and a high angular distribution of the radiation. An undulator from a magnet array as depicted here produces a narrow distribution of wavelength and angular divergence. The rest of the setup is comparable to the laboratory setup, albeit the intensity of the radiation is orders of magnitude higher, which allows for finer collimation slits and longer collimation distances and SDDs.}
\label{fig:synchrotronSAXS}
\end{figure}

While the setup in general is similar to that of a laboratory setup there are some key differences between a synchroton and a laboratory SAXS setup. Most of the differences are based on radio protection needs and are therefore immaterial to this description in terms of the SAXS measurement itself. The other main difference is in the production of the x-rays itself. Current setups at synchrotrons use undulators in order to periodically accelerate charged particles (usually electrons/positrons) perpendicular to the direction of propagation of the particle beam. This creates a very brilliant, nearly perfectly monochromatic x-ray beam along the direction of the electron beam. The monochromaticity can further be improved by a monochromator crystal. Fig.\ref{fig:synchrotronSAXS} shows an example of an synchrotron SAXS setup. After that, the collimation is very similar to that of a laboratory SAXS setup, only the materials are chosen to be thicker in most cases to improve the absorption characteristics. Due to the monochromaticity the brilliance, coherence and signal-to-noise ratio are significantly better than that of a laboratory SAXS setup, since there is no bremsstrahlung spectrum to contribute to the background. In terms of achievable wavelength there is no limitation to use a specific K-$\alpha$ line of any specific material. Often common wavelengths are chosen to better correspond to laboratory measurements on identical samples. One option that is also available in some synchrotrons is the tunability of the wavelength in order to measure resonance effects in the atomic structure of the sample (anomalous SAXS, \emph{ASAXS})\cite{haubold1999characterization} or better chose the accessible $Q$ space. Tab.\ref{tab:synchrotronSAXS} summarizes some of the performance figures of current synchrotron SAXS setups. For most synchrotron SAXS beamlines beam damage, especially for organic samples, is an issue and has to be taken into account when planning an experiment.

\begin{table}
\centering
\begin{tabular}{l|l}

Parameter & value \\ 
\hline 
\hline
SDD & 0.8-4 m \\ 
\hline 
Pixel resolution & 172$\times$172$\mu$m \\ 
\hline 
Flux & $10^{18}$ photons s$^{-1}$ \\ 
\hline 
wavelength $\lambda$ & 0.54 - 1.38 \AA \\ 
\end{tabular}

\caption{Performance parameters for a state of the art synchrotron SAXS beamline, here P03 at DESY.\cite{roth2011situ}}

\label{tab:synchrotronSAXS}

\end{table}

\section{SANS setups}
\index{SANS}
In contrast to x-rays, sufficient numbers of free neutrons can only be obtained by nuclear processes, such as fission, fusion and spallation. As large-scale facilities are needed to create the processes at a suitable rate to perform scattering experiments with them, the only facilities where neutron scattering today can be performed is at fission reactor sources and spallation sources. This of course also leads to larger efforts in terms of biological shielding.

It is an inherent feature of those reactions that the reaction products show a wide distribution of energies, with peak energies ranging up to 3 MeV kinetic energy per neutron. This leads to deBroglie wavelengths in the femtometer region, which is unsuitable for SANS scattering experiments. Thus, in order to obtain a coherent beam it is not only necessary to collimate the neutrons but also to moderate and monochromatize them. Both processes result in losses in usable flux, since the phase space of neutrons cannot be compressed by lenses, as is the case for photons.

The moderation process is performed by collision processes in a moderator medium. The moderator is a material at temperatures around 25 K or below and the resulting neutron spectrum is a Maxwell-Boltzmann spectrum of the corresponding temperature. This results in peak wavelengths around 4 \AA\, for the neutron beam. Neutron scattering instruments can be run both in time-of-flight mode or monochromatic mode.

\begin{figure}
\centering
\begin{minipage}[l]{0.75\textwidth}
\Large{a)}

\includegraphics[width=\textwidth]{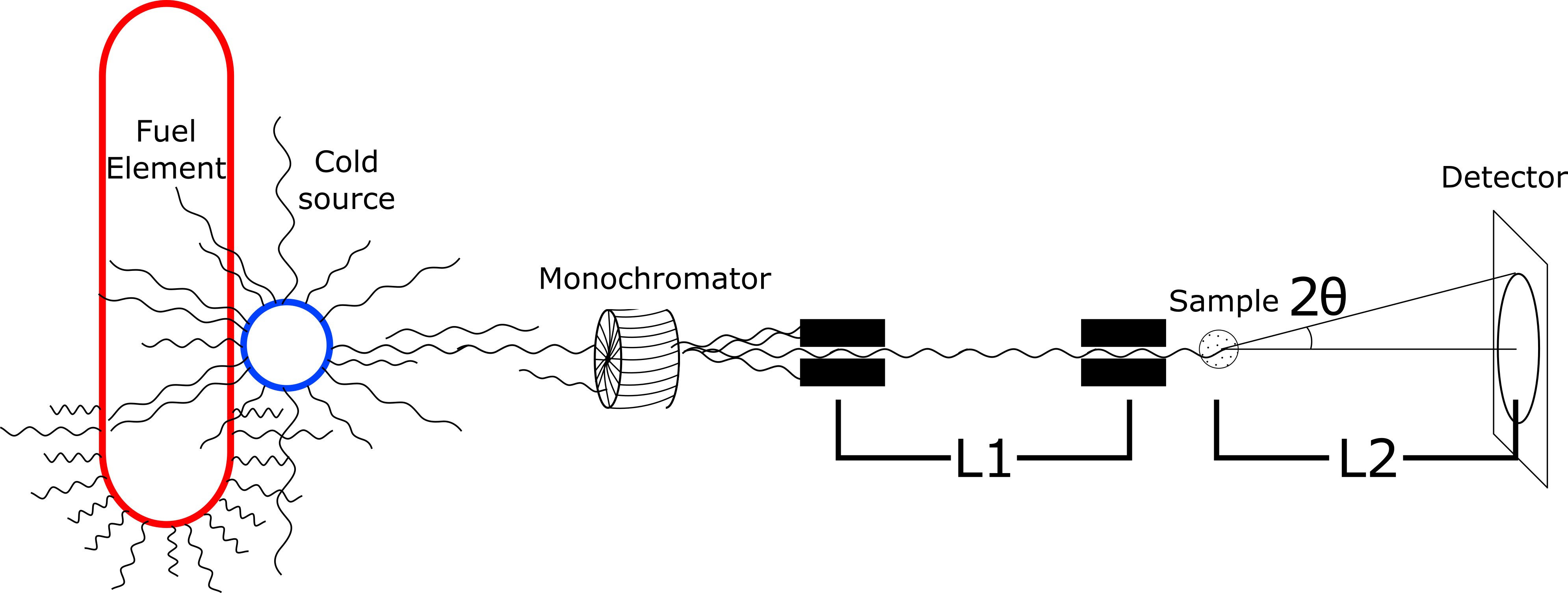}
\end{minipage}

\begin{minipage}[l]{0.75\textwidth}
\Large{b)}

\includegraphics[width=\textwidth]{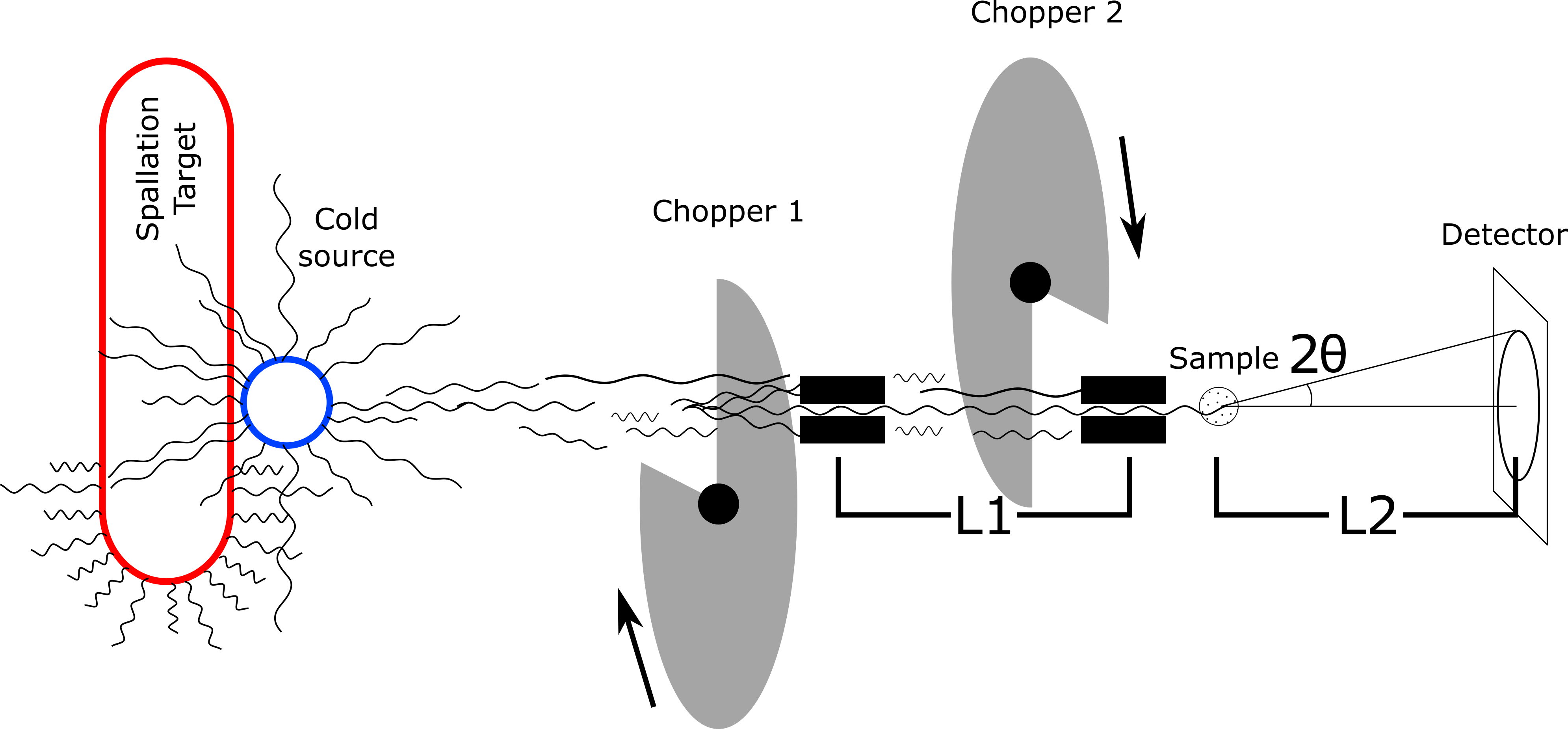}
\end{minipage}
\caption{a) Continuous source SANS setup and b) pulsed source SANS Setup. In both cases the neutron source (red) creates hot neutrons of a short wavelength. A cold source (blue) vessel (usually filled with cold $^2$H or $^2$D) is moderating the neutrons down to slower speeds, i.e. longer wavelengths. In both cases the collimation distance and SDD is widely adjustable for most instruments, with lengths between 1 m up to 30 m. In a SANS instrument at a continuous source a monochromator (a turbine with slightly inclined channels) selects a certain wavelength (usually between 3 and 15 \AA) and afterwards the setup is very much like the one shown for SAXS setups, except that the whole instrument is larger. In case of a pulsed source, choppers (rotating discs with transparent openings for neutrons) define a start and an end time for each pulse. Since neutrons, different from x-rays, are particle waves, their wavelength determines their speed. Thus, the wavelength is determined by measuring the time of arrival at the detector for each neutron. For an optimized neutron transport all components are usually evacuated.}
\label{fig:SANSsetup}
\end{figure}

A schematic of a SANS instrument is shown in Fig.\ref{fig:SANSsetup}. Both cases with a monochromator and a chopper setup for time-of-flight are presented. In a continous source the neutron flux has to be interrupted for the timing of time-of-flight mode while for pulsed sources there is an inherent interruption of the neutron flux.

This moderation and collimation process in consequence means that neutrons always show an, albeit small, distribution of wavelengths and therefore a lower signal to noise level than x-ray sources. Spin and isotopic incoherence add to that. Beam damage however is nigh on impossible with the weakly interacting neutrons.

\newpage
\section{Indirect space and Small-Angle Scattering}
The need for the resolution of small angles can be directly derived from Bragg's equation

\begin{equation}
n \lambda = 2d\cdot\sin\theta
\label{eq:bragg}
\end{equation}

with $n$ being the order of the diffraction, $d$ being the distance between two scatterers, $2\theta$ as the scattering angle and $\lambda$ the wavelength of the incoming beam. In order to get interference the incoming beam has to have a wavelength that corresponds to the investigated size regime, which in both cases is on the order of a few Angstroms. Using Bragg's equation with $n=1$, $d=50$\AA\, and $\lambda = 1$\AA\, we arrive at $0.01 = \sin\theta \approx \theta$. Thus, the largest structures to be resolved are determined by the smallest achievable angle.

\begin{figure}
\centering
\includegraphics[width=0.75\textwidth]{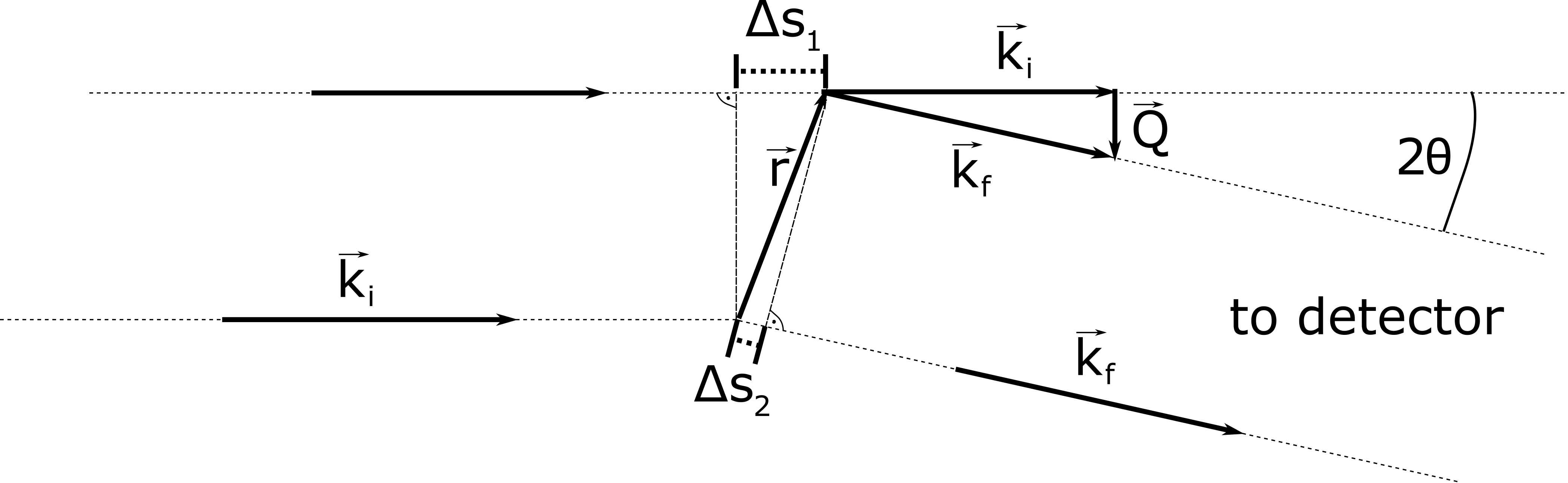}
\caption{Construction of $\vec{Q}$. The incoming and final wavevectors $\vec{k}_i$ and $\vec{k}_f$ define both the scattering vector $\vec{Q}$ as well as the path length difference $\delta=\Delta s_1 -\Delta s_2$. Here it is important to note that the selection of the center of origin is arbitrary and thus can be chosen to be at the center of the construction. The calculation of the length of $\vec{Q}$ is then given by Eq.\ref{eq:Q}.}
\label{fig:Qconstruction}
\end{figure}

In order to allow for a setup and wavelength independent data evaluation the data is recorded in terms of $Q$ or indirect space. The construction of that $Q$ space from two scattering points is shown in Fig.\ref{fig:Qconstruction}. From that the magnitude of $\vec{Q}$, which here for simplicity is $|\vec{Q} |= Q$, can be derived as

\begin{equation}
Q=\frac{4\pi}{\lambda}\sin\theta.
\label{eq:Q}
\end{equation}

Even though $\vec{Q}$ is strictly speaking a vector, for most small angle problems only the absolute value $Q$ is of interest, hence this simplification is reasonable. This is due to the isotropic scattering picture of a majority of small-angle scattering data. Another simplification that is often used is the small-angle approximation for the sine with $\sin\theta = \theta$, which is very well valid for small angles. Combining Eqs.\ref{eq:bragg} and \ref{eq:Q} also delivers a useful expression for the approximation of inter-particle distances or correlation lengths

\begin{equation}
d = \frac{2\pi}{Q}.
\label{eq:distances}
\end{equation}

\section{Resolution limits}

SAS is working based on the interference of coherent radiation. That in itself imposes some limitations on the samples and properties that can be investigated. Any loss of coherence will limit the resolution.

In terms of size, the path length difference due to the scattering at the object has to be of the same order of magnitude as the wavelength of the incoming radiation, analogous to light interference at a double slit. Concerning the analysis in indirect space, also the limited size of the detector and coherence volume of the sample has to be taken into account. 

\subsection{Coherence resolution limit}
Coherence can be measured along the beam propagation, where different speeds of neutrons can mean that after a certain flight time they are not in the same wave package anymore. Essentially this longitudinal coherence is a measure of the wavelength measured in its own spread, i.e.

\begin{equation}
L_{coh} = \frac{\lambda ^2}{\Delta \lambda} = \lambda \left(\frac{\Delta \lambda}{\lambda}\right) ^{-1}
\label{eq:coherence}
\end{equation}

This essentially measures, how many multiples of the wavelength a neutron can propagate, before the error is large enough to make up a full wavelength. With neutrons often a $\Delta \lambda / \lambda = 0.1$ is achieved in instrumentation, which means that the longitudinal coherence length $L_{coh}$ is approximately 10 times the wavelength, so on the order of 3-10 nm. This is not a hard upper resolution limit, but it limits the path length difference created by scattering events. If the path length differences are too large then this will not be visible. Using Bragg's equation \ref{eq:bragg} we can calculate the angle belonging to a certain path length difference. Assuming a 10 nm structure and a 20$^\circ$ angle we arrive at approximately 7 nm path length difference. This means, for $7\AA = 0.7 $nm wavelength neutrons this will not affect them in a neutron scattering instrument. However, this is usually limited by the angular Q-resolution as shown in the next section.

This also explains why small-angle scattering is so powerful with relatively large objects. In equation \ref{eq:bragg} the size of the object and the sine of the angle are a product. If the object becomes large, with a smaller angle you can compensate and still fulfil the conditions for coherent interference scattering.

\subsection{Smallest and largest structure}
From Eq.\ref{eq:distances} it is apparent that to measure small distances in the sample structure, large $Q$ are required and vice versa. The large Q-values in the case of SAS are limited by the size of the detector, nuclear distances are usually measured in diffractometer instruments which cover a larger solid angle.

Large detectors for a SAS instrument would cover an angle of +/- $20^\circ$. This results in $Q=0.7 \AA ^{-1}$ for a wavelength of 3 $\AA$ and hence a distance of approximately 9 $\AA$ for the smallest resolvable structure. In order to reach a much larger path length difference, where the coherence limit would impact the measurement larger detectors and measurements of smaller distances would be necessary, which are outside the experimental scope for SAS experiments.

More interesting is the lower angular limit, so the largest observable structure. This is for a lot of SANS instruments on the order of 0.1$^\circ$ or even smaller. This means, with the same wavelength as above of 3 $\AA$ this results in a maximum observable size of 300 nm at $Q\approx 3\cdot 10^{-3} \AA^{-1}$. If this can be reduced farther, where some instruments can reach on the order of $Q\approx 10^{-4} \AA^{-1}$ which results in an observable size up to a micrometer in the extreme limit.

\section{Fourier Transform and Phase Problem}
\label{section:phaseProblem}

Considering the spacing of only two scattering centers as in the last section needs to be extended to an arrangement of scattering centers for evaluation of macroscopic samples, where each atom/molecule can contribute to the scattered intensity. Since the incoming wave at location $\vec{x}$ can be considered to be an plane wave it can be described by

\begin{equation}
\psi(\vec{x},t)
=
A_{\mathrm{inc}}
\exp\left[
i(\vec{k}\cdot\vec{x}-\omega t)
\right],
\label{eq:amp}
\end{equation}

where $\psi$ is the wave amplitude, $A_{\mathrm{inc}}$ the incoming amplitude, $\omega$ the angular frequency, and $\vec{k}$ the incoming wave vector [1].

In order to calculate the phase shift $\Delta\phi$ between waves scattered from two centers, as shown in Fig.~\ref{fig:Qconstruction}, we need the difference in travelled distance $\delta$ between the two waves. This yields

\begin{equation}
\Delta\phi
=
\frac{2\pi\delta}{\lambda}
=
\vec{Q}\cdot\vec{r},
\end{equation}

where the scattering vector is defined as

\begin{equation}
\vec{Q}=\vec{k}'-\vec{k}.
\end{equation}

This then leaves us with the wave scattered by the first scattering center

\begin{equation}
A_1(\vec{x},t)
=
A_{\mathrm{inc}}\,b\,
\exp\left[
i(\vec{k}'\cdot\vec{x}-\omega t)
\right],
\end{equation}

where $b$ is the scattering length, and the corresponding scattered wave from the second scattering center

\begin{equation}
A_2(\vec{x},t)
=
A_1(\vec{x},t)\exp(i\Delta\phi).
\end{equation}

Here the wave number is defined as $|\vec{k}| = \frac{2\pi}{\lambda}$.

Therefore

\begin{equation}
A_2(\vec{x},t)
=
A_{\mathrm{inc}}\,b\,
\exp\left[
i(\vec{k}'\cdot\vec{x}-\omega t)
\right]
\exp(i\vec{Q}\cdot\vec{r}).
\end{equation}

This can then be combined into the full description of the amplitude with both contributions to

\begin{eqnarray}
A(\vec{x},t)
&=&
A_1(\vec{x},t)+A_2(\vec{x},t)
\\
&=&
A_{\mathrm{inc}}\,b\,
\exp\left[
i(\vec{k}'\cdot\vec{x}-\omega t)
\right]
\left(
1+\exp(i\vec{Q}\cdot\vec{r})
\right).
\label{eq:fullAmp}
\end{eqnarray}

note that the scattering efficiency $b$ for each scattering center will later be discussed for both x-rays and neutrons.

Since only intensity can be observed at the detector, we need to consider the modulus square, calculated with the complex conjugate of the expression itself

\begin{eqnarray}
I(\vec{Q})
&=&
\frac{1}{V_{\mathrm{sample}}}
A(\vec{x},t)A^*(\vec{x},t)
\\
&=&
\frac{A_{\mathrm{inc}}^2 b^2}{V_{\mathrm{sample}}}
\left(
1+\exp(i\vec{Q}\cdot\vec{r})
\right)
\left(
1+\exp(-i\vec{Q}\cdot\vec{r})
\right).
\label{eq:Amp2Int}
\end{eqnarray}

Here the time and absolute location dependencies in Eq.\ref{eq:fullAmp} have cancelled each other out, so we can neglect them and are left with a function that solely depends on the scattering vector $\vec{Q}$ and the relative position of the particles $\vec{r}$ with respect to each other. The normalisation to the scattering volume $V_{\mathrm{sample}}$ is due to the fact, that the scattered intensity is directly proportional to the scattering volume. Utilising those dependencies allows us to generalize Eq. \ref{eq:fullAmp} to the case of $N$ identical scattering centers with

\begin{equation}
A(\vec{Q})
=
A_{\mathrm{inc}}
\sum_{i=1}^{N}
b_i
\exp(i\vec{Q}\cdot\vec{r}_i).
\label{eq:ampSLD}
\end{equation}

For identical scatterers with constant scattering length $b$, this simplifies to

\begin{equation}
A(\vec{Q})
=
A_{\mathrm{inc}}\,b
\sum_{i=1}^{N}
\exp(i\vec{Q}\cdot\vec{r}_i).
\end{equation}

The $\vec{r}_i$ here signify the relative locations of all scattering centers in the sample, relative to either simply the first scattering center or any arbitrary center chosen. Indeed the choice of coordinate origin is arbitrary and any choice is mathematically identical. Replacing the sum by a weighed integral allows also the calculation for the case of a (quasi)continuous sample with number density $\rho(\vec{r})$:

\begin{equation}
A(\vec{Q})
=
A_{\mathrm{inc}}\,b
\int_{V_{\mathrm{sample}}}
\rho(\vec{r})
\exp(i\vec{Q}\cdot\vec{r})
\,\mathrm{d}V.
\label{eq:AmplitudeIntegral}
\end{equation}

This is the Fourier transform of the number density of scattering centers with identical scattering efficiency $b$, it can also be applied for numerous scattering efficiencies, if a space dependent scattering power (scattering length density) $\rho_b(\vec{r}) = b(\vec{r})\rho(\vec{r})$ is used under the integral.

In the following, the incoming amplitude is normalized to unity,$A_{\mathrm{inc}}=1$, such that all intensities are given relative to the incoming beam intensity.

However, since the phase information got lost while obtaining the intensity as an absolute square of the amplitudes, there is no direct analytic way of performing an inverse Fourier transform. This is why this is called the phase problem.\index{Phase problem} Also, as described above, in a wide range of cases it is enough to investigate the modulus of $\vec{Q}$, neglecting its vector nature.

\section{Scattering Efficiency}
Since the physical scattering event is very dissimilar for x-rays and neutrons they shall be discussed separately here. However, it should be noted, that the nature of the scattering process does not impact on the method of data evaluation in general. Only in very specific cases, such as contrast matching or polarized scattering there is any discernible difference.

\subsection{Scattering with x-rays}
X-rays, as photons, interact with the sample via electromagnetic interaction. For the purpose of this manuscript it is sufficient to note that the vast majority only interact with the electron shell around the atoms and thus effectively map the electron density within the sample. Interactions with the nucleus would only occur at very high energies, which are not usually used in elastic scattering. In a rough approximation the strength of the electromagnetic interaction scales with $Z^2$, meaning that heavy elements, such as a wide range of common metals, scatter considerably stronger than light ones, like hydrocarbon compounds. For element analyses there is also the possibility of resonance scattering, where the chosen x-ray energies are close to the resonance gaps in the absorption spectrum of specific elements (ASAXS).\cite{haubold1999characterization} 

Based on Thomson scattering the scattered intensity at angle $2\theta$ is

\begin{eqnarray}
I(2\theta)&=&I_0 \left(\frac{e^2}{mc^2}\right)\frac{1+\cos^2 2\theta}{2}\\
\frac{I}{I_0}&=&\left(\frac{\d\sigma}{\d\Omega}\right)_2 = r^2 _e \frac{1+\cos^2 2\theta}{2}
\label{eq:Thomson}
\end{eqnarray}

Here we also introduced the differential scattering cross section $\frac{d\sigma}{d\Omega}$ for a single electron and $r_e$ being the radius of an electron. This means that the total probability for a scattering event to occur into a solid angle $d\Omega$ is exactly that value for a single, isolated electron. This probability is in units of an area. Thus, the scattering length for a single electron $b_e$ is defined as the square root of that:

\begin{equation}
b_e=r_e\sqrt{\frac{1+\cos ^2 2\theta}{2}}
\label{eq:elektronSLD}
\end{equation}

With those previous equations it is again important to note that small-angle scattering is mainly concerned with small angles, thus that $\cos 2\theta \approx 1$ is a very good approximation. This is also, together with backscattering, the location of the highest intensity and negligible polarization effects. The numeric values for the constants used here are $r_e=2.818\times 10^{-15}$ m and the scattering cross section for a single electron $\sigma_e=6.65\times10^{-29} \mbox{ m}^2\mbox{ }= 0.665 $ barn after integration over the full solid angle. As apparent with integration over the full solid angle, the relation is $\sigma = 4\pi b_e^2$.

Since usually the goal is to find the distribution of scattering centers in a volume, the density of scattering length per unit volume is of interest. This is the scattering length density (SLD)

\begin{equation}
\rho_b (\vec{r}) = \sum_i b_i N_i = \int_V b(\vec r) dV.
\label{eq:SLD}
\end{equation}

Here $b_i$ is the scattering length of each individual scattering center. For the second equality in this equation we did a transition from discrete scattering centers to a continuous spatial distribution of scattering length in the scattering volume $V$.

A very common way of expressing scattering efficiency is using electron units. As can be seen in Eq.\ref{eq:ampSLD} the scattering amplitude is only determined by the 
scattering length
of a single electron apart from the Fourier transform of the local density. This means the scattering intensity in electronic units can be expressed as

\begin{equation}
I_{eu}(Q)=\frac{I(Q)}{I_0b_e^2}
\end{equation}

This means, with appropriate calibration, if there is an intensity of $I_{eu} = 200\,b_e^2$ at a certain $Q$, that the size scale corresponding to that $Q$ vector has 200 electrons per unit volume. 

Since photons interact mainly with the electron shell, there is also an angle dependency accounting for the time averaged location probability of the electrons in the shell, which may or may not be spherical, depending on the electronic configuration of that specific atom. This would then lead to a SLD in terms of $b_e(Q)=b_e f(Q)$ with $f$ being the atomic scattering factor for any specific element. This important to take note of, when there is a structure or form factor on the same size scale as a single atomic distance $Q=\frac{2\pi}{1.54\mbox{ \scriptsize{\AA}}} = 4.08\mbox{ \AA}^{-1}$. This is usually not in the regime of interest for small-angle scattering and will mostly vanish in the incoherent background.

Another incoherent background effect is Compton scattering, where inelastic processes change the wavelength during the scattering process. This is however again strongly suppressed at small angles. The wavelength shift occurring based on Compton scattering is following this expression

\begin{equation}
\Delta\lambda = \frac{h}{mc}2 \sin^2\theta
\end{equation}

The prefactor is $\frac{h}{mc} = 0.02426$ \AA. It is also obvious that at large angles $2\theta = 180$\deg\ the energy transfer is maximal. Since we are always investigating angles close to $\theta = 0$ the wavelength shift and hence the incoherent background is negligible compared to other experimental factors, such as slits and windows scattering.

\subsection{Scattering with neutrons}

\subsubsection{Scattering length}
While we could rely on classical electromagnetic scattering theory for x-rays, for neutrons we have to apply quantum mechanics for the calculation. While for photons we could rely on the relatively long-range electromagnetic interaction of the electron shell there is no simple corresponding expression for neutrons, since the neutron always experiences the potential of the nucleus directly. This means, the potential experienced is both directly influenced by the weight of the core, i.e. number of nuclei, which corresponds to the depth of the potential, as well as radius of the core, which corresponds to the width of the potential. Keep in mind, the same calculation could be done for an electromagnetic potential and the result would be the Thomson scattering formula \ref{eq:Thomson}. Similar approaches to the one shown here were shown by Hammouda or Tong \cite{hammouda1995tutorial, ScatteringTheory}. For additional details, see Appendix \ref{SchroedingerAppendix}.

In order to calculate the scattering length density, we have to start with the time-independent Schrödinger equation

\begin{equation}
-\frac{\hbar ^2}{2 m}\nabla ^2\psi + V \psi = E \psi.
\label{eq:Schrodinger}
\end{equation}

With $k_1^2 =\frac{2m(E+V_0)}{\hbar ^2} $ we can write a solution for this for weak scattering potentials $V_0$ at low scattering energies $E$, which is exactly the case for neutron scattering. The solution for the scattering length $a_s = b$ incorporates the radius of the scattering nucleus $a$:

\begin{equation}
a_s = b = a - \frac{\tan (k_1 a)}{k_1}.
\end{equation}

\begin{figure}
\centering
\includegraphics[width=0.75\textwidth]{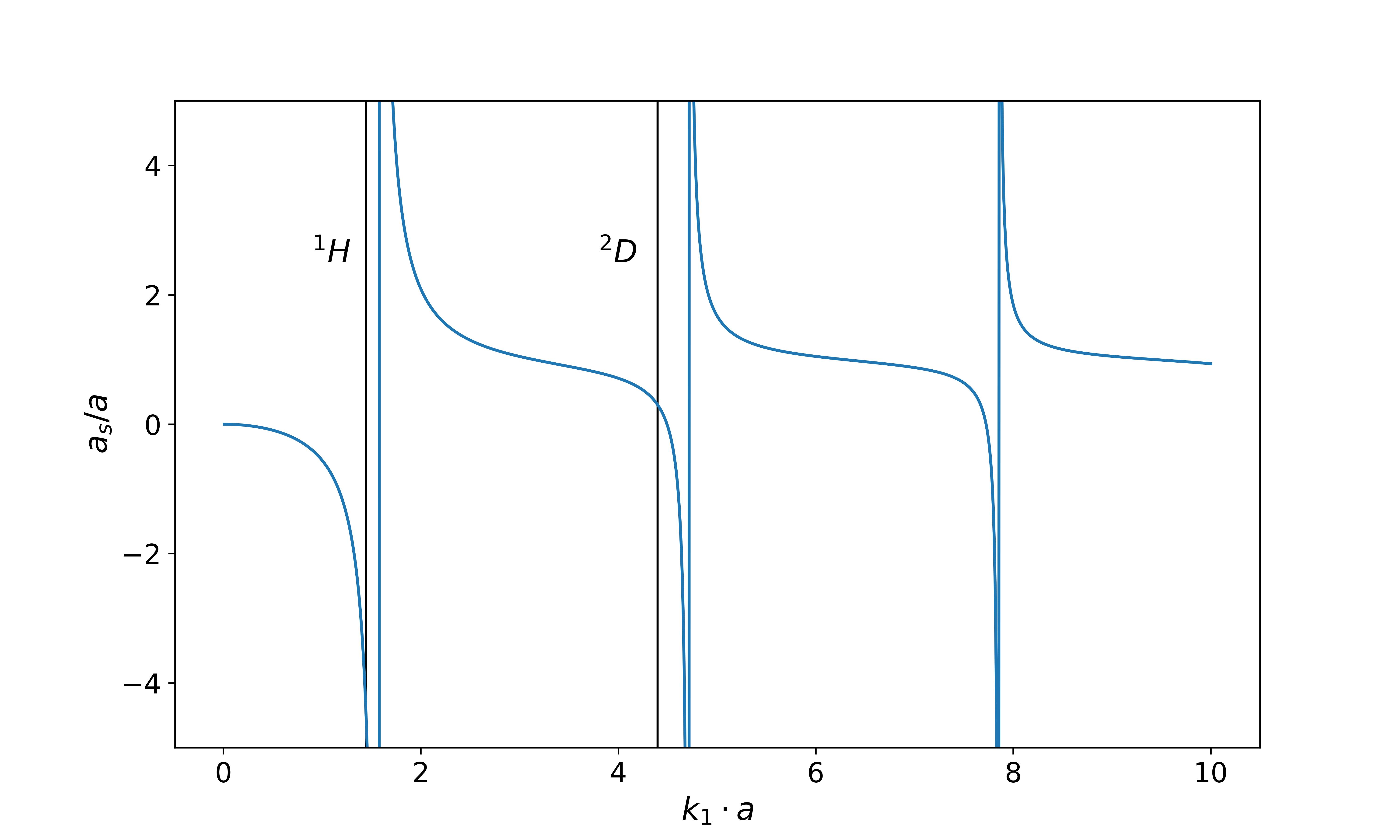}
\caption{Behavior of scattering length $b=a_s$ as a function of nuclear radius $a$ and $k_1$. Using representative values for example for hydrogen ($a= 8.5 \cdot 10^{-16}$ m) and deuterium ($a= 2.1 \cdot 10^{-15}$ m) it becomes apparent how those strong differences in scattering length happen, and also why this has an especially strong impact for light elements, where the radius strongly changes by adding or removing one nucleon. Values for hydrogen and deuterium are marked. As a side note, this can also be used to calculate the depth of the nuclear potential via neutron scattering, since the energy $E$ of the neutron is known and the only free parameter is $V_0$.}
\label{fig:NeutronCrossSection}
\end{figure}

A representation of this is given in Fig. \ref{fig:NeutronCrossSection}. The strong variation due to minute changes in the numerical value make it clear, why tabulated values are used in most cases. In addition, until here we only considered spherically symmetric s-wave scattering. Since magnetic momenta are not spatially uniform a lot of the approximations used above would not work in that case.

Based on that we usually rely on tabulated values for the cross sections and scattering lengths of different elements and isotopes (see Tab.\ref{tab:neutronSLD}) and can then write the cross section and scattering length relation as

\begin{table}
\centering
\begin{tabular}{c|c}
Element & scattering length $b_{coh}$/$10^{-14} m$ \\ 
\hline 
\hline
$^1$H  & -0.374 \\ 
\hline 
$^2$D & 0.667 \\ 
\hline 
C & 0.665 \\ 
\hline 
N & 0.936 \\ 
\hline 
O & 0.580 \\ 
\hline 
Si & 0.415 \\ 
\hline 
Br & 0.680\\
\end{tabular} 
\caption{Coherent scattering length of several elements and isotopes.}
\label{tab:neutronSLD}
\end{table}

\begin{equation}
\frac{d\sigma}{d\Omega}=b^2
\end{equation}

One important correlation between the two, if taking a naive approach to the view of a sphere that is hit from one side, which in projection is a circle, is that the total scattering cross section is given by the surface of a sphere with radius $b$. This counterintuitive fact can be understood as a quantum mechanical particle not only interacting with the projected surface, but with the whole surface of the potential:

\begin{equation}
\sigma=4\pi b^2
\end{equation}

\begin{figure}
\centering
\includegraphics[width=0.75\textwidth]{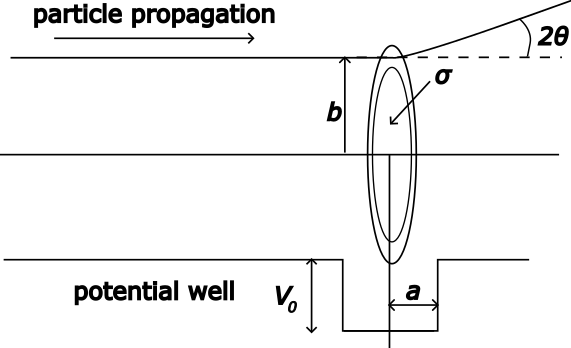}
\caption{Sketch of scattering length $b$, scattering cross section $\sigma$, potential depth $V_0$ and radius $a$ for a particle scattering event. Please note that $\sigma$, which could be seen as a circle here is quantum mechanically rather a sphere.}
\label{fig:CrossSectionSketch}
\end{figure}

The general correlation between the different entities is shown in a sketch in Fig.\ref{fig:CrossSectionSketch}

That said, only coherent scattering can form interference patterns, i.e. no change of the nature of the radiation can take place during the scattering process. However, since the neutron can change its spin orientation through spin-spin coupling during the scattering process that may happen, depending on the spin orientation of the sample nuclei. Those are completely statistical processes.

As neutrons are fermions, which have spin $1/2$ the possible outcomes after a scattering process with a nucleus of spin $i$ are $i+1/2$ and $i-1/2$, and the associated possible spin states are

\begin{eqnarray}
\mbox{number of states } i+1/2&:&2(i+1/2)+1=2i+2\\
\mbox{number of states } i-1/2&:&2(i-1/2)+1=2i\\
\mbox{total number of states }&:&4i+2.
\label{eq:spinStates}
\end{eqnarray}

This immediately shows, that only for the case $i=0$ there can be only two states. Since it is impossible to know the spin state of non-zero spin nuclei under ambient conditions, the differential cross section becomes a two-body problem of the form:

\begin{equation}
\frac{d\sigma}{d\Omega} = \sum _{i,j} \left\langle b_i b_j\right\rangle \exp-i\vec{Q}(\vec{r_i}-\vec{r_j})
\end{equation}

Here $\left\langle b_i b_j\right\rangle$ is the expectation value of the product of scattering lengths for each $b_ib_j$ combination possible given isotope and spin variability. For this there is only one coherent outcome, where $b_i = b_j$, which then results in

\begin{equation}
\left\langle b_i b_i\right\rangle = \left\langle b_i ^2\right\rangle = \left\langle b ^2\right\rangle.
\end{equation}

All other cases result in $b_i \neq b_j$ and therefore

\begin{equation}
\left\langle b_i b_j\right\rangle_{i\neq j} = \left\langle b_i \right\rangle \left\langle b_j\right\rangle = \left\langle b\right\rangle^2.
\end{equation}

This then results in 

\begin{equation}
\frac{d\sigma}{d\Omega} =  \langle b^2\rangle \cdot \sum_{j,k} \exp\left(-i\vec{Q}(\vec{r_i}-\vec{r_j})\right) +N(\langle b^2\rangle-\langle b\rangle^2).
\end{equation}

Here $\sqrt{\langle b^2\rangle} = b_{coh}$ signifies the coherent scattering length, since it contains information about the structure of the sample via $\vec{r_{ij}}$ and $\sqrt{\langle b^2\rangle-\langle b\rangle^2} = b_{inc}$ is the incoherent scattering length not allowing for coherent constructive interference. This cannot be suppressed instrumentally; therefore, isotopes with low incoherent scattering length are often chosen in neutron scattering to suppress the incoherent background. Both coherent and incoherent scattering lengths can be used separately together with Eq.\ref{eq:SLD} to obtain the corresponding scattering length densities.

\begin{figure}
\centering
\includegraphics[width=0.75\textwidth]{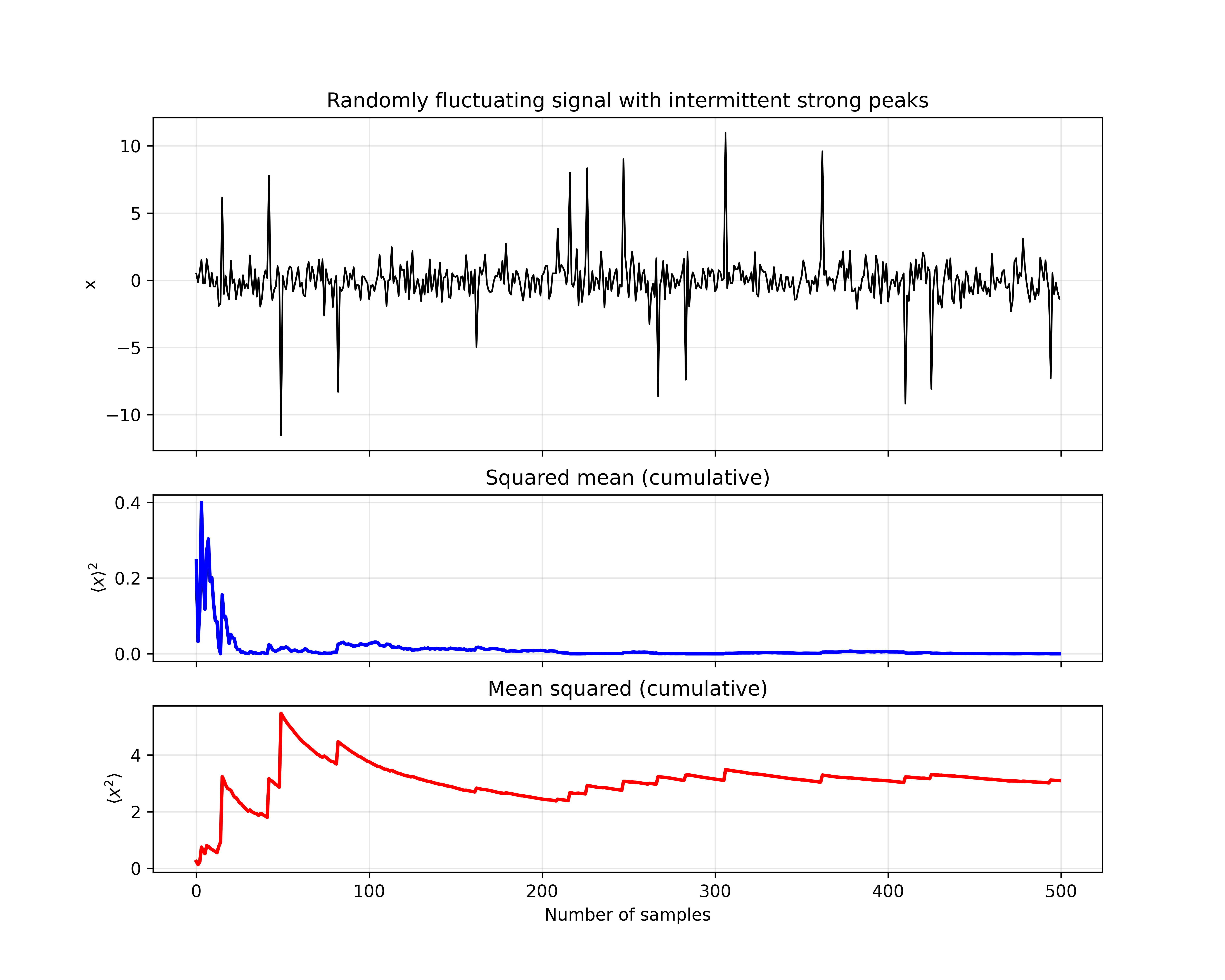}
\caption{Illustration on the difference between squared mean and mean squared to show the idea of autocorrelation for similar scattering centers in mean squared and the idea of summing over non-identical scattering centers for squared mean.}
\label{fig:MeanSquaredSquaredMean}
\end{figure}

For the calculations above it is useful to recall the difference between the squared mean $\langle x\rangle ^2$ and the mean squared $\langle x^2\rangle$. Mean squared squares every value, sums it up and then divides by the number of values, which therefore is a combination of autocorrelation and fluctuations around zero, while squared mean measures the coherent average overall. They are connected by the variance $\langle x^2\rangle - \langle x \rangle ^2 = Var(x)$, which makes it immediately apparent than a strongly fluctuating sample (in this case a lot of different scattering centers) result in a high variance, low autocorrelation (with identical scatterers) and a large incoherent background. This can also be seen with the example of a simple statistical white noise signal (see Fig. \ref{fig:MeanSquaredSquaredMean}).

\subsection{Scattering Cross Section and Contrast Matching}
\begin{figure}
\centering
\includegraphics[width=0.75\textwidth]{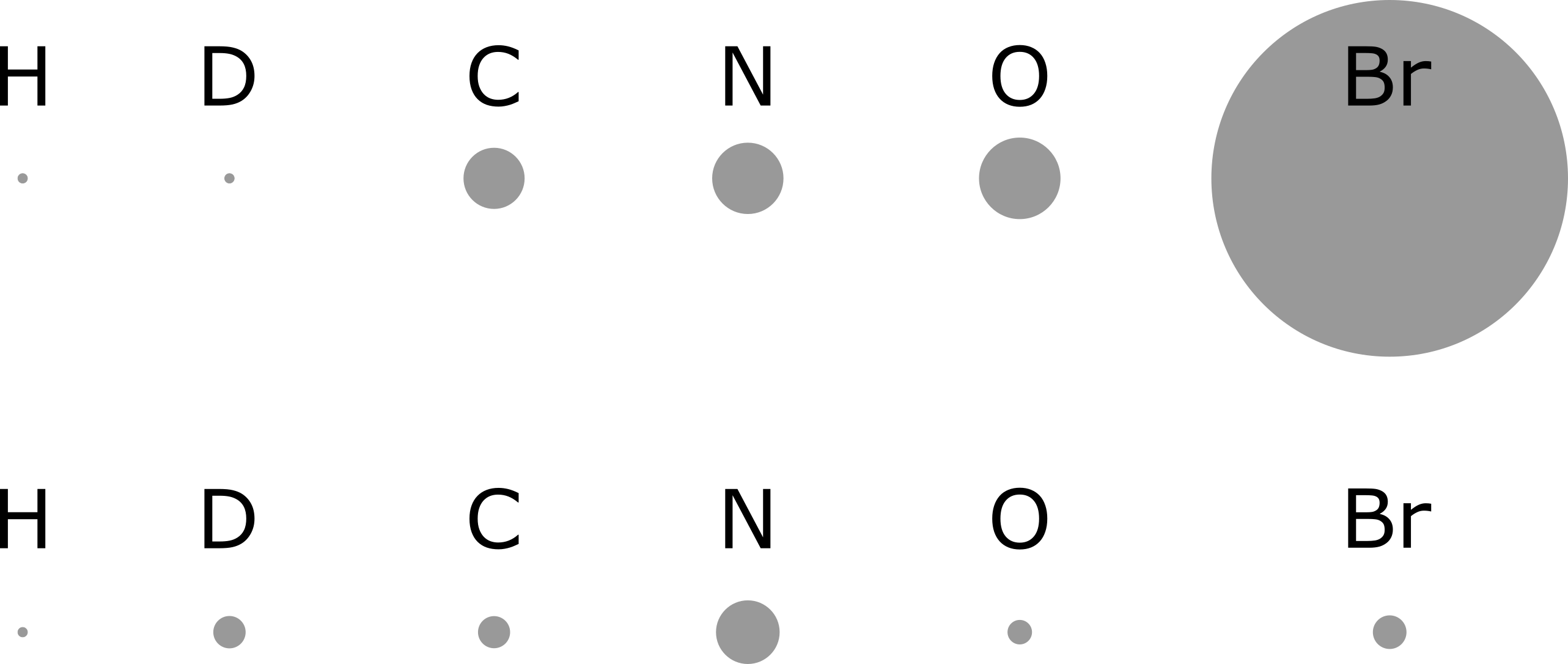}
\caption{Coherent cross-sections for selected elements for x-rays (top) and neutrons (bottom). The coherent scattering cross section scales linearly with the diameter of the circles. It is apparent, that the $Z^2$ dependency strongly emphasizes heavy elements in x-ray scattering, whereas for neutrons even single isotopes can be distinguished. However, for neutrons there is no simple analytic expression for the scattering cross-sections.}
\label{fig:CrossSection}
\end{figure}

As described above there is a $Z^2$ dependency of the cross section of atoms in case of x-rays and the cross section values for neutrons are taken from tabulated values. The resulting differences in cross section are illustrated in Fig.\ref{fig:CrossSection}. Because different isotopes have very different cross sections for neutron scattering, in some cases it is possible to replace certain isotopes in order to arrive at desired contrast conditions.

\begin{figure}
\centering
\includegraphics[width=0.75\textwidth]{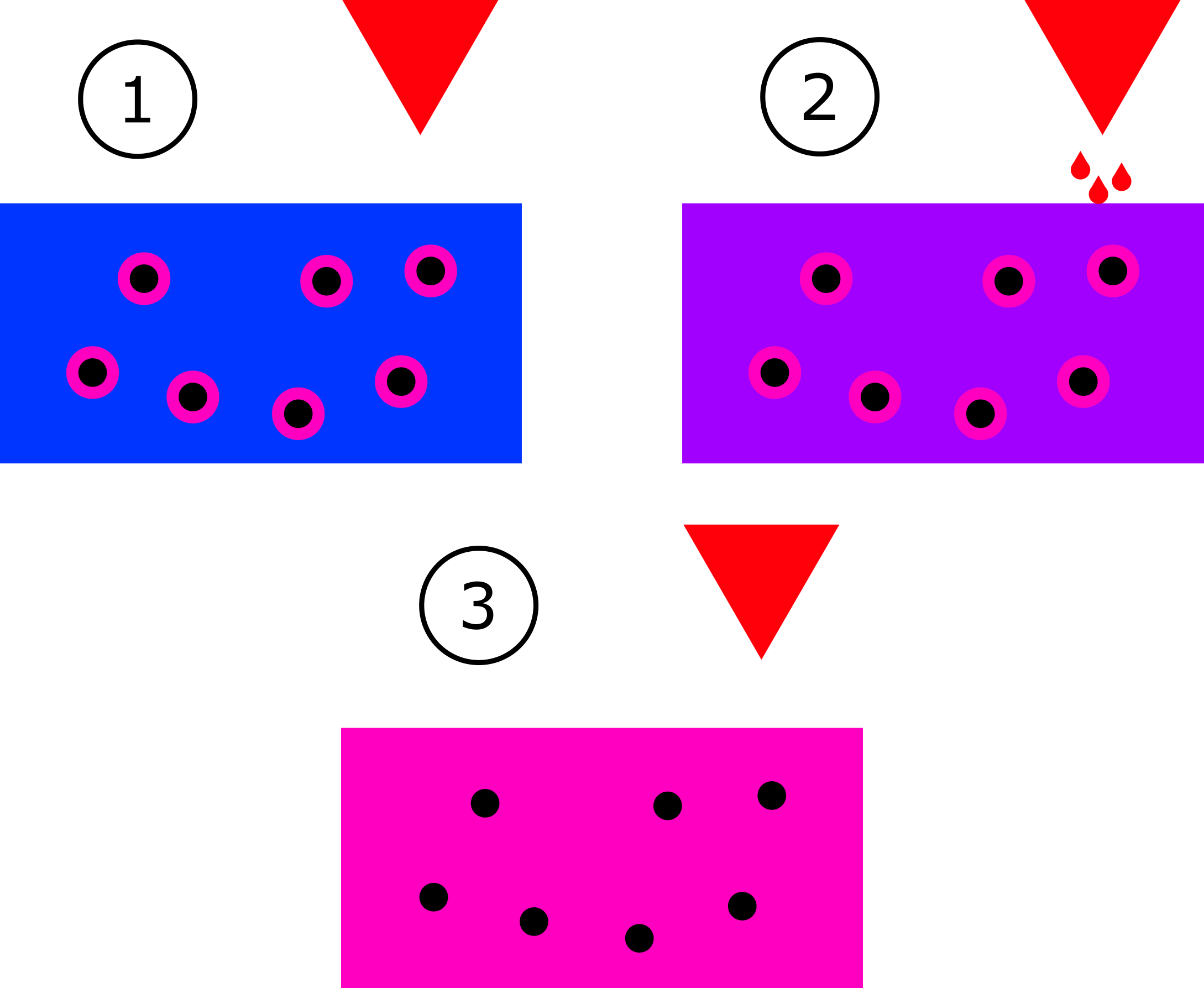}
\caption{Illustration of the concept of contrast matching. In step \ding{172} there are micelles with a corona (pink) dissolved in a solution (blue). The scattering length density of the corona is between the SLD of the solvent and its deuterated counterpart (red). In step \ding{173} the deuterated solvent is added to the solution, which changes the contrast conditions. Finally, in step \ding{174} a sufficient amount of deuterated solvent has been added, so the contrast between the corona and the solvent has vanished. Now the micellar cores can be measured directly.}
\label{fig:ContrastMatchConcept}
\end{figure}

One of the most important examples for that technique, called contrast matching, is replacing hydrogen by deuterium. This leaves the chemical composition of the sample unchanged, and hydrogen is extremely abundant in most organic compounds. The concept can in some cases be extended to be used as the Babinet principle, in order to suppress background scattering, since it is extremely preferable to have a solvent with a low background and a solute with a higher background than vice versa. A sketch of the concept is shown in Fig.\ref{fig:ContrastMatchConcept}.

This method allows highlighting otherwise hidden features of the sample or suppressing dominant scattering in order to better determine a structure with a lower volume fraction and therefore less scattering contribution. Examples for that application are highlighting the shell of a sphere, by matching the core or vice versa. Also for protein samples certain structures can be matched, so that only distinct features are visible.

\begin{figure}
\centering
\includegraphics[width=0.75\textwidth]{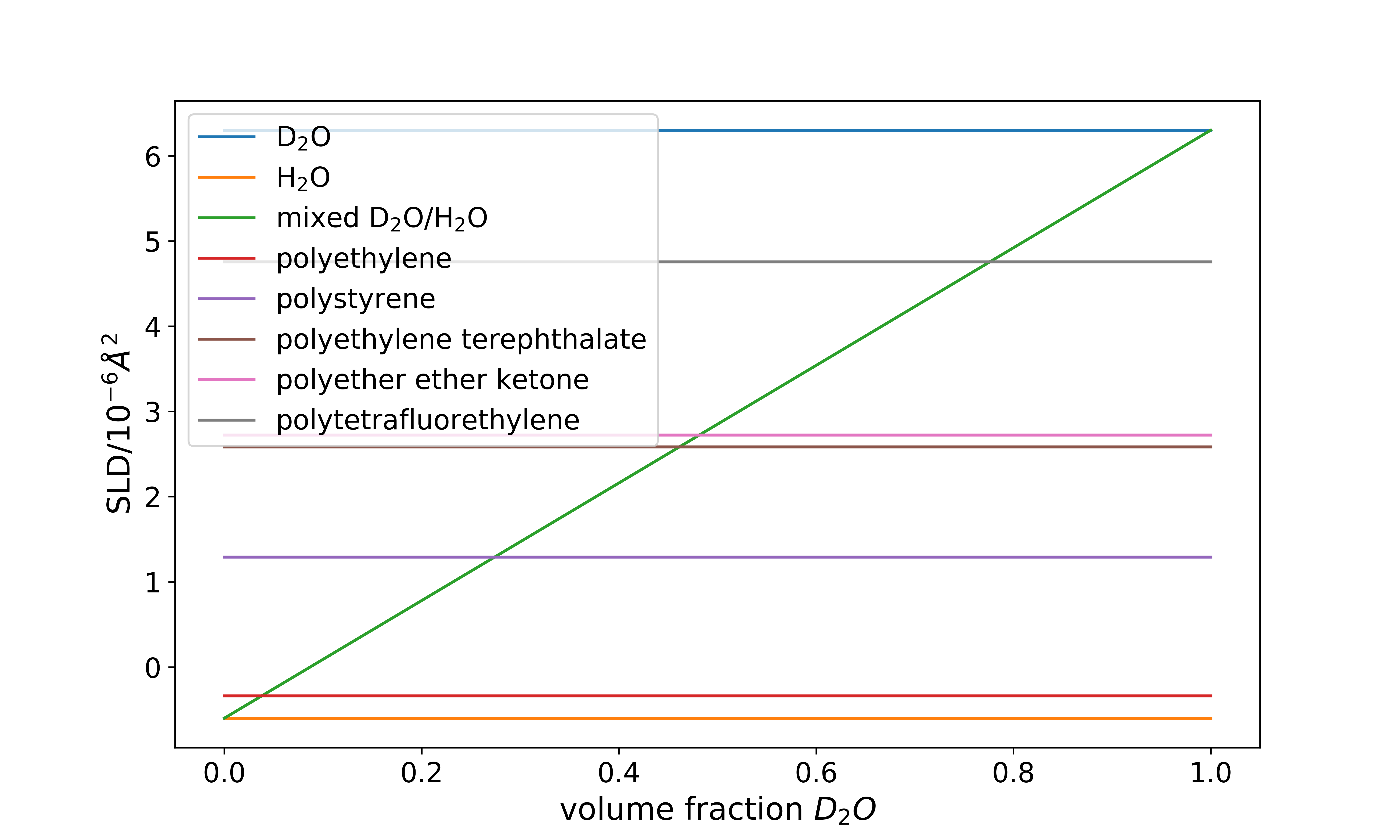}
\caption{Semi-analytic way to determine the necessary solvent deuteration for contrast matching. The concentration at the matching point, where the solvent has the same SLD as the polymer particles, is determined by the crossing of the mixed D$_2$O/H$_2$O SLD line and the SLD line of the respective polymer. For the calculation the scattering length density of water is calculated to -0.6$\cdot10^{-6}$\AA$^2$ and the SLD of heavy water is calculated to 6.3$\cdot10^{-6}$\AA$^2$.}
\label{fig:ContrastMatchSLD}
\end{figure}

In order to apply contrast matching, mostly the solvent is changed. In some rare cases also the polymer or other sample is synthesized with a different isotope composition. Here the finding of the correct H/D fraction of the solvent shall be shown. Fig.\ref{fig:ContrastMatchSLD} gives an example of how to find the correct H/D fraction in a semi-analytic way. The underlying principle is expressed by

\begin{eqnarray}
\rho_{b,sample} &=& \rho_{b, H_2O}\times H + \rho_{b, D_2O}\times D\\
H&\equiv & 1\\
D&=&\frac{\rho_{b,sample}-\rho_{b, H_2O}}{\rho_{b, D_2O}}.
\end{eqnarray}

This way the volume of heavy water for each unit volume of protonated (usual) water can be calculated. It is also apparent from that calculation that only mixtures with a scattering length density between water and heavy water can be matched, and that the equations above only cover the non-trivial cases, where pure water or heavy water is not suitable. The actual volumes can then be calculated with $V_{\mbox{water}}=\frac{H}{H+D}$ and $V_{\mbox{heavy water}}=\frac{D}{H+D}$.

\begin{figure}
\centering
\includegraphics[width=0.75\textwidth]{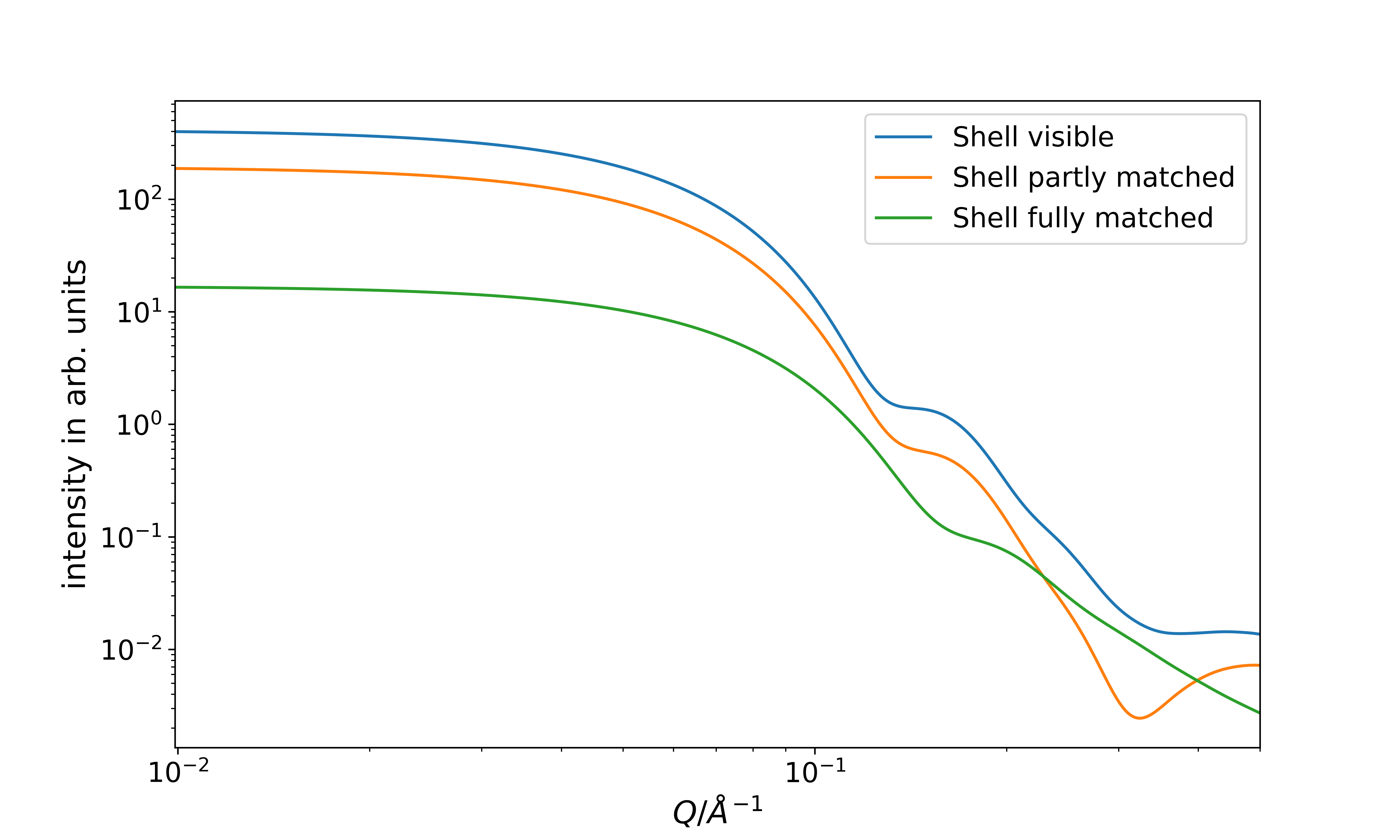}
\caption{Scattering curves for micelles with unmatched, partially matched and completely matched corona. The curves correspond to the scenarios \ding{172}, \ding{173} and \ding{174} in Fig.\ref{fig:ContrastMatchConcept}. Here two effects can be observed. The corona is only 50\% of the radius of the core, hence it influences the scattered intensity at higher angles than the core itself, the scattering feature at $Q$=0.15\AA$^{-1}$ corresponding to the micellar core is therefore quite stable, while the intensity at higher $Q$ changes drastically. Considering the forward scattering the dependence of the scattering contrast between solvent and core is directly visible. The matched out corona shows the least contrast, and therefore the lowest forward scattering intensity, while the unmatched corona has the highest contrast and the highest intensity. This approach is also used, when an analytic approach to find the matching D$_2$O/H$_2$O concentration cannot be found. Several concentrations are tested and where a minimum in the scattered intensity is found, the contrast can be assumed to be matched.}
\label{fig:ContrastMatchCurves}
\end{figure}

A prominent example for contrast matching is the matching out of the shell or core of a micelle. The contrast behavior and the resulting scattering curves are shown in Fig.\ref{fig:ContrastMatchCurves}. Essentially contrast matching can improve the fitting procedure, if well known parts of the structure are matched out or emphasized by the contrast matching. This then delivers two or more different data sets that all should return comparable results. Another option is the reconstruction of embedded particles in a larger structure. Also here, the overall fitting procedure can benefit from two fits with mutually corroborating results.

One concept that shall also be mentioned here is magnetic (spin-) contrast. In this context Fig.\ref{fig:ContrastMatchConcept} can be understood to be particles with a magnetic shell. As long as the spins are not aligned there is no contrast between the shell and the solvent (step \ding{174}). When an external magnetic field aligns the spins in the shell, a contrast between the shell and the solvent emerges (\ding{172}). Several other possibilities with and without polarization analysis are possible, however that is beyond the scope of this manuscript.

\section{Form factors}
\label{sec:formFactors}
\index{Form factor}
There are different usages of the term form factor in the literature. The term applied in SAS is the squared magnitude of the Fourier transform of the scattering length density contrast of an individual particle, describing scattering from particles of a given shape and size. Note this is structurally similar, but not identical to the atomic form factor in X-ray crystallography, where the form factor is the Fourier transform of the electron density around an atom and therefore represents a scattering amplitude. The scattered intensity is obtained from the squared magnitude of this quantity. 

\begin{figure}
\centering
\includegraphics[width=0.75\textwidth]{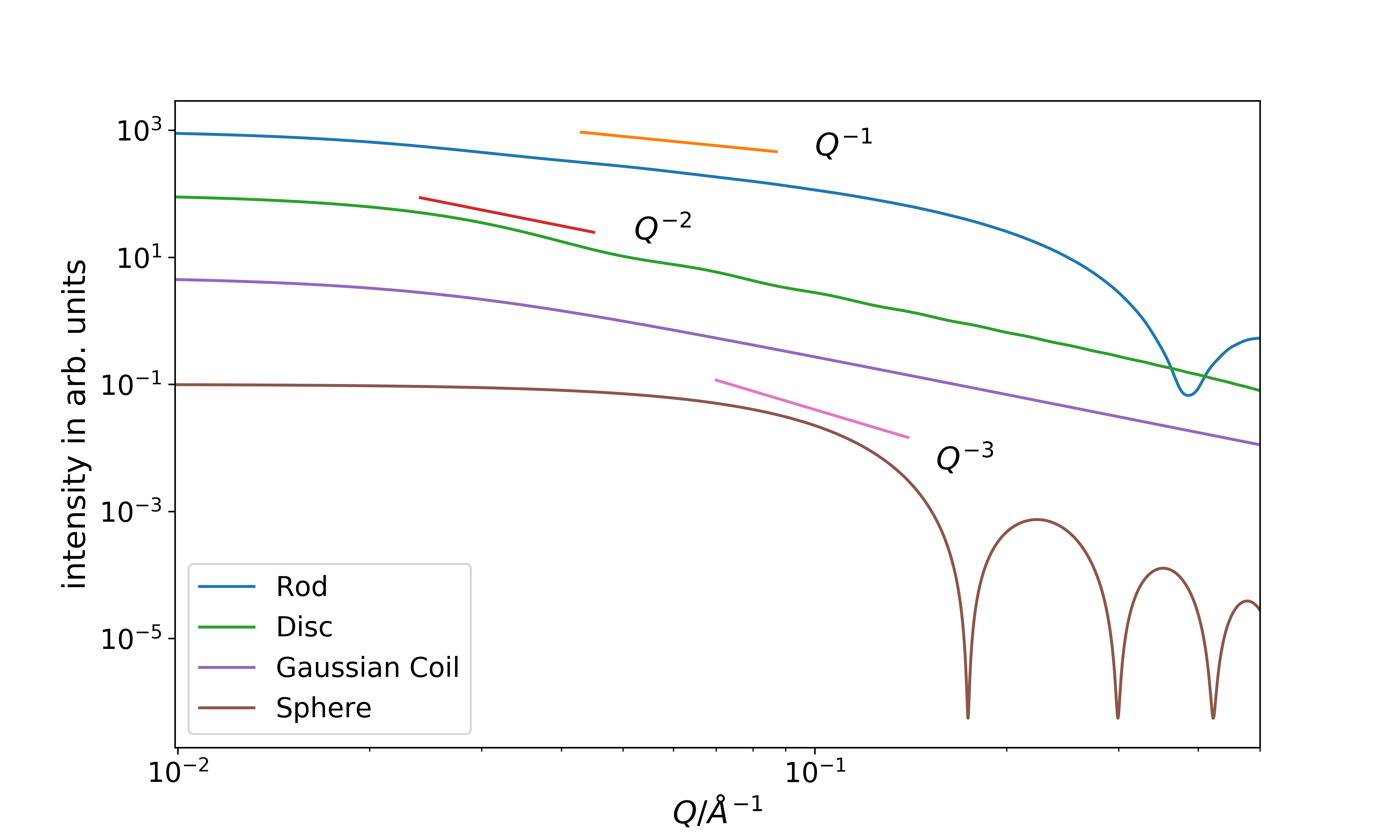}
\caption{Form factors for several scattering geometries. The slopes at the onset of the form factor after the plateau are shown, which is mostly determined by the fractal dimension of the scattering object. Here it also becomes apparent that solely relying on that slope may lead to misinterpretation between similarly scaling objects, here Gaussian coils and discs.}
\label{fig:formFactors}
\end{figure}

As described above, the phase problem usually prevents an analytic reconstruction of the structure from the scattered intensity by an inverse Fourier transform. There are approaches attempting the direct reconstruction of direct space information \cite{weyerich1999small} or reconstruction from bead model annealing / Monte Carlo simulation \cite{koutsioubas2016denfert,grant2018ab}. All these approaches have in common that a direct analytic expression for the scattering is not foreseen, and therefore it is hard to use them as a starting point of the analysis. In the past, the model based analysis has been the most applied approach for the analysis of small-angle scattering data. Here, predetermined structures undergo a Fourier transform, whose result is then used to calculate a scattering pattern. This results in the most cases in analytic expressions that can be directly fitted to the data and are often used in a reference-based manner in order to determine the structure of the sample. As most geometric forms can be approximated either as a sphere, a disk or a rod (see Fig.\ref{fig:formFactors}) these forms are discussed below. More elaborate structures are available and can in principle be calculated for any structure where the form can be described by an analytic expression. A short, and by no means complete, list of programs for the evaluation of SAS data is SasView (https://www.sasview.org), SasFit (https://kur.web.psi.ch/sans1/SANSSoft/sasfit.html) and Scatter (http://www.esrf.eu/UsersAndScience/Experiments/\newline
CRG/BM26/SaxsWaxs/DataAnalysis/Scatter\#).

The mathematical rationalization for the use of form (and actually structure) factors comes from Eqs. \ref{eq:Amp2Int} and \ref{eq:ampSLD}. With those two equations we can write

\begin{eqnarray}
I(\mathbf{Q}) & = & \frac{1}{V_{\mathrm{sample}}}\left\langle A(\mathbf{Q})A^*(\mathbf{Q}) \right\rangle \\
		& = &
\frac{b^2}{V_{\mathrm{sample}}}
\left\langle
\left\{
\sum_i ^{N_{\mathrm{scatterers}}}\exp(-i\mathbf{Q}\cdot\mathbf{r}_i)
\right\}
\left\{
\sum_j ^{N_{\mathrm{scatterers}}}\exp(i\mathbf{Q}\cdot\mathbf{r}_j)
\right\}
\right\rangle .
\label{eq:doubleSumFF}
\end{eqnarray}

Here we are running the sums over all scattering centers ${N_{\mathrm{scatterers}}}$ in the sample. Exploiting the properties of the sum and the exponential function we can pull this together in one term of the form

\begin{equation}
I(\mathbf{Q})
=
\frac{b^2}{V_{\mathrm{sample}}}
\left\langle
\sum_i ^{N_{\mathrm{scatterers}}} \sum_j ^{N_{\mathrm{scatterers}}} \exp[-i\mathbf{Q}\cdot(\mathbf{r}_j-\mathbf{r}_i)]
\right\rangle .
\end{equation}

This expression means that we sum over all scattering centers at positions $\vec{r}_i$ with identical scattering power $b$. Since the expression is independent of the choice of coordinate origin, we may separate the sum into contributions originating from scattering centers within the same particle and from scattering centers belonging to different particles.

For an assembly of $N$ identical particles and using $\vec{r}_{ji}=\vec{r}_j-\vec{r}_i$, this becomes

\begin{eqnarray}
I(\mathbf{Q}) & = &\frac{N N_{\mathrm{scatterers}} b^2}{V_{\mathrm{sample}}} \biggl\langle \frac{1}{N_{\mathrm{scatterers}}} \left\{ \sum_{i,j \,\mathrm{in\,same\,particle} }^{N_{\mathrm{scatterers}}} \exp(-i\mathbf{Q}\cdot\mathbf{r}_{ji}) \right\} \label{eq:formStructureFactorA}\\
 & \times &\left\{ 1+ \frac{1}{N} \sum_{k,l \,\mathrm{centers\,of\,different\,particles} }^{N} \exp(-i\mathbf{Q}\cdot\mathbf{r}_{kl}) \right\} \biggr\rangle .
\label{eq:formStructureFactorB}
\end{eqnarray}

Two things happened here: In Eq. \ref{eq:formStructureFactorA}, since all particles are identical, we only need to perform the internal sum once and multiply it by $N$, since this is the number of particles. Also, we need to be careful about ${N_{\mathrm{scatterers}}}$. While it denoted all scatterering centers in a sample before, here it only denotes the scattering centers belonging to one particle. We normalize by the number of scattering centers, so that at $Q=0$ this whole section becomes 1. You may find in other textbooks this normalization is missing, if the scatterers are considered in a continuum (see \ref{eq:AmplitudeIntegral}). Then this normalization is included into the density of scattering centers in space. Note for the following sections, if only \ref{eq:formStructureFactorA} is considered it cancels immediately out, so this is only necessary if both intra- and inter-particle relations are considered. 

In Eq. \ref{eq:formStructureFactorB} we define $\vec{r}_{kl}=\vec{r}_k-\vec{r}_l$ as the distance between the centers of different, but identical, scattering particles. So here the sum is running over all particles, instead of the scattering centers within the particle. We obtain the term $1$ from the self-correlation contribution with $\vec{r}_{kl}=0$, for which $\exp(0)=1$. Furthermore, the second term is divided by $N$ in order to avoid counting inter-particle distances multiple times.

Here we can include a simple mental check: In Eq. \ref{eq:doubleSumFF} we ran over each particle twice, once in the left sum, once in the right sum. In Eqs. \ref{eq:formStructureFactorA} and \ref{eq:formStructureFactorB} we also run over each particle twice, once inside the particle and again for each particle that is there. Also, looking at Eq. \ref{eq:formStructureFactorB} we can see the Fourier transform of the location of particles. If the particles have a strong correlation the Fourier transform will return a non-zero value for some $\vec Q$-values, imagine the Fourier transform of a sine-wave where you get two Dirac $\delta$ functions for the frequency of the sine wave either side of zero. If the particles show no correlation at all, you have a Fourier transform of an average constant distribution, which in turn is a sine of cosine wave with some amplitude. In the limit of many particles, the division by $N$ means the second term essentially goes to zero, so you are only left with the leading $1$, which gives you exactly the form factor and no additional contributions from correlation between the particles, as is expected if there is no correlation between the particles.

Until this point no approximation has been made; we merely reordered the sums according to the particle arrangement inside the scattering volume. For the sake of simplicity we assume from here on identical particles, no coupling between intra- and inter-particle correlations, and particles that are orientationally averaged as indicated by $\left\langle A(\vec{Q})A^*(\vec{Q})\right\rangle$.

This allows us to write the scattered intensity from a collection of identical particles in the form

\begin{eqnarray}
I(\mathbf{Q}) &=& \frac{N N_{\mathrm{scatterers}} A_0^2}{V_{\mathrm{sample}}}F(Q)S(Q)
\\
&=& \frac{N N_{\mathrm{scatterers}} \Delta\rho_b^2 V^2}{V_{\mathrm{sample}}}F(Q)S(Q)
\\
&=& \rho N_{\mathrm{scatterers}} \Delta\rho_b^2 V^2 F(Q)S(Q),
\\
& \textrm{with} &
\nonumber
\\ \rho &=& \frac{N}{V_{\mathrm{sample}}},
\\
A_0 &=& A(Q=0) = \Delta\rho_b V,
\\
f(\mathbf{Q}) &=& \frac{1}{\sqrt{N_{\mathrm{scatterers}}}} \sum_{\mathrm{i\,in\,same\,particle}}^{N_{\mathrm{scatterers}}} \exp \left( -i \vec{Q} \vec{r}_i\right),
\\
F(Q) &=& |f(\mathbf{Q})|^2 = \frac{1}{N_{\mathrm{scatterers}}} \sum_{\mathrm{i,j\,in\,same\,particle}}^{N_{\mathrm{scatterers}}} \exp \left( -i \vec{Q} \vec{r}_{ji}\right),
\\
& \textrm{and} &
\nonumber
\\
S(\mathbf{Q}) &=& 1+ \frac{1}{N} \sum_{k,l \,\mathrm{centers\,of\,different\,particles}}^N
\exp(-i\mathbf{Q}\cdot\mathbf{r}_{kl}) .
\end{eqnarray}

Here, $\rho$ is the particle number density, $V$ the particle volume, and $\Delta\rho_b$ the scattering length density contrast. The forward scattering amplitude is given by

\begin{equation}
A_0  =  A(Q=0) = \Delta\rho_b V.
\end{equation}

Please note here that $A_0$ is the amplitude for a complete particle, not for individual scattering centers.

This defines the form factor $F(\vec{Q})$ and structure factor $S(\vec{Q})$ as dimensionless descriptions of the particle shape and the inter-particle correlations, respectively. The incoming beam intensity was normalized by setting $A_{\mathrm{inc}}=1$ in section \ref{section:phaseProblem}.

\subsection{Sphere}
\index{Form factor!Sphere}
The analytic expression for the scattering created by identical spheres of radius $R$ with a number density of $\rho$ is

\begin{equation}
I(Q)
=
\rho
\left[
3V\Delta\rho_b
\frac{\sin(QR)-QR\cos(QR)}{(QR)^3}
\right]^2
=
\rho \Delta\rho_b^2 V^2 F(Q),
\end{equation}

with $V$ the volume of a single sphere, and $\Delta\rho_b=\rho_{b,\mathrm{sphere}}-\rho_{b,\mathrm{solvent}}$ the SLD contrast between the sphere and the solvent.

\begin{figure}
\centering
\includegraphics[width=0.65\textwidth]{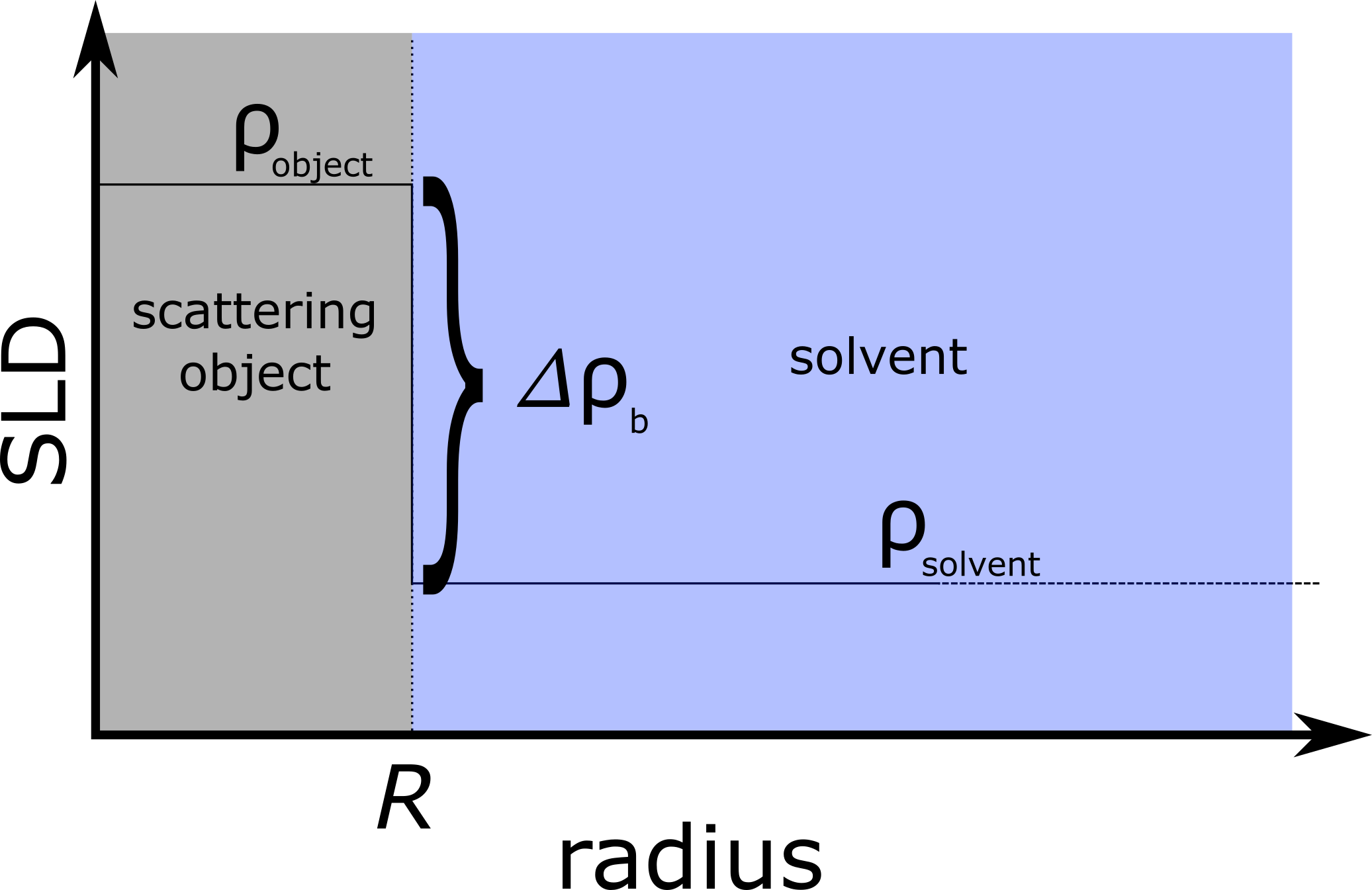}
\caption{Depiction of the SLD distribution along the radius of a sphere. $\rho _0$ is the SLD contrast, i.e. the SLD difference between the scattering particle and the solvent. $R$ is the radius of the sphere.}
\label{fig:SLDSphere}
\end{figure}

Following the convention that $f(Q)$ is the form factor amplitude and $F(Q)=|f(Q)|^2$ the corresponding form factor amplitude is

\begin{equation}
f(Q)
=
3
\frac{\sin(QR)-QR\cos(QR)}{(QR)^3},
\label{eq:SphereForm}
\end{equation}

and therefore

\begin{equation}
F(Q)=|f(Q)|^2.
\end{equation}

This expression can be reached by using a SLD description like a step function as depicted in Fig.\ref{fig:SLDSphere}. As a sphere is already spherically symmetric this can be directly put into the Fourier transform

\begin{eqnarray}
f(\vec{Q})
&=&
\frac{1}{V\Delta\rho_b}
\mathcal{F}(\rho_b(\vec{r}))
\\
&=&
\frac{1}{V\Delta\rho_b}
\int_V
\rho_b(\vec{r})
\exp\left(
-i\vec{Q}\cdot\vec{r}
\right)
\,dV
\label{eq:GeneralFT}
\\
&=&
\frac{1}{V\Delta\rho_b}
\int_{\phi=0}^{2\pi}
\int_{\theta=0}^{\pi}
\int_{r=0}^{R}
\rho_b(\vec{r})
\exp\left(
-i\vec{Q}\cdot\vec{r}
\right)
r^2\sin\theta
\,dr\,d\theta\,d\phi
\\
&=&
\frac{1}{V\Delta\rho_b}
\int_{\phi=0}^{2\pi}
\int_{\theta=0}^{\pi}
\int_{r=0}^{R}
\rho_b(r)
\exp\left(
-iQr\cos\theta
\right)
r^2\sin\theta
\,dr\,d\theta\,d\phi
\label{eq:scalarProduct}
\\
&=&
\frac{1}{V\Delta\rho_b}
\int_{\phi=0}^{2\pi}
\int_{u=-1}^{1}
\int_{r=0}^{R}
\rho_b(r)
\exp\left(
-iQru
\right)
r^2
\,dr\,du\,d\phi
\label{eq:substitute}
\\
&=&
\frac{4\pi}{V\Delta\rho_b}
\int_{r=0}^{R}
\rho_b(r)
\left(
\frac{\sin(Qr)}{Qr}
\right)
r^2
\,dr
\\
&=&
\frac{4\pi}{V}
\int_{r=0}^{R}
\frac{\sin(Qr)}{Qr}
r^2
\,dr
\\
&=&
\frac{4\pi}{V}
\frac{
\sin(QR)-QR\cos(QR)
}{
Q^3
}
\\
&=&
3
\frac{
\sin(QR)-QR\cos(QR)
}{
R^3Q^3
}
\\
&=&
3
\frac{
\sin(QR)-QR\cos(QR)
}{
(QR)^3
}.
\end{eqnarray}

Here Eq.~\ref{eq:scalarProduct} used the identity $\vec{Q}\cdot\vec{r}=Qr\cos\theta$ with $\theta$ being the enclosed angle and in Eq.~\ref{eq:substitute} $\cos \theta$ was replaced by $u$. In addition, spherical symmetry was exploited for the integration over the solid angle.

This corresponds exactly to the squared term in Eq.~\ref{eq:SphereForm}, which is nothing else than the squared form factor amplitude calculated here. As this is only the scattering for a single isolated sphere, the number density needs to be included to reflect the absolute scattered intensity. In case of neutron scattering this is the case for most instruments. X-ray instruments are often not calibrated to absolute scattering intensities and therefore require an arbitrary scaling factor. Similar approaches can be used for other analytic representations of form factors.

\subsection{Cylinder}
\index{Form factor!Rod}
In order to derive the same form factor for a cylinder we assume a cylinder with
\begin{itemize}
\item radius $R$
\item length $L$
\item the symmetry axis running along $z$ for cylindrical coordinates
\item uniform scattering length density $\Delta\rho_b$
\end{itemize}

and we separate the coordinate along the symmetry axis and the coordinate perpendicular to the symmetry axis with

\begin{eqnarray}
q_z & = & q \cos \alpha \\
q_{\perp} & = & q \sin \alpha \\
& \Rightarrow & \\
\vec{Q}\cdot\vec{r} & = & q_{\perp}r\cos\phi+q_z z
\end{eqnarray}

with $\alpha$ being the angle between the $Q$-vector and the cylinder symmetry axis and $\phi$ being the coordinate angle. Here the scattering vector is defined as $\vec{Q}=\vec{k}'-\vec{k}$ [1].

With that

\begin{eqnarray}
f(\vec{Q})
&=&
\frac{1}{V\Delta\rho_b}
\mathcal{F}(\rho_b(\vec{r}))
\\
&=&
\frac{1}{V}
\int_V
\exp\left(
-i\vec{Q}\cdot\vec{r}
\right)
\,dV
\\
&=&
\frac{1}{V}
\int_{-L/2}^{L/2}
\int_0^{2\pi}
\int_0^R
\exp\left(
-i(q_{\perp}r\cos\phi+q_z z)
\right)
r
\,dr\,d\phi\,dz
\\
&=&
\frac{1}{V}
\int_{-L/2}^{L/2}
\exp(-iq_z z)
\,dz
\int_0^{2\pi}
\int_0^R
\exp(-iq_{\perp}r\cos\phi)
r
\,dr\,d\phi .
\end{eqnarray}

Now we consider the longitudinal part

\begin{equation}
\int_{-L/2}^{L/2}
\exp(-iq_z z)
\,dz
=
L
\frac{
\sin(q_zL/2)
}{
q_zL/2
}.
\end{equation}

and the radial part separately.

\begin{equation}
\int_0^{2\pi}
\exp(-iq_{\perp}r\cos\phi)
\,d\phi
=
2\pi J_0(q_{\perp}r)
\end{equation}

creates a zero-th order Bessel function from the angular integral, leaving only the radial coordinate for integration. This leads to

\begin{equation}
2\pi
\int_0^R
rJ_0(q_{\perp}r)
\,dr
=
2\pi R^2
\frac{
J_1(q_{\perp}R)
}{
q_{\perp}R
}.
\end{equation}

Note here that we used that the order of the Bessel-functions can be increased by integration.

The longitudinal and angular part can now be combined by drawing out the cylinder volume $V=\pi R^2 L$ from the constants in both results, resulting in

\begin{equation}
f(Q,\alpha)
=
2
\frac{
\sin\left(
\frac{1}{2}QL\cos\alpha
\right)
}{
\frac{1}{2}QL\cos\alpha
}
\frac{
J_1(QR\sin\alpha)
}{
QR\sin\alpha
}.
\end{equation}

Following the convention that $F(Q)=|f(Q)|^2$, the corresponding intensity form factor is

\begin{equation}
F(Q,\alpha)=|f(Q,\alpha)|^2.
\end{equation}

Here we assume the cylinders are very thin, i.e. $R\ll L$ and therefore $Q R \ll 1$. This results in the approximation

\begin{equation}
2\frac{J_1(x)}{x}\approx1
\end{equation}

eliminating the second part of the formfactor. With that

\begin{eqnarray}
F(\vec{Q},\alpha)
&=&
|f(Q,\alpha)|^2
\\
&=&
\left[
\frac{
\sin\left(
\frac{1}{2}QL\cos\alpha
\right)
}{
\frac{1}{2}QL\cos\alpha
}
\right]^2 .
\end{eqnarray}

For randomly oriented cylinders we now perform the orientational average over $\alpha$

\begin{eqnarray}
F(\vec{Q})
&=&
\int_0^{\pi/2}
F(Q,\alpha)
\sin\alpha
\,d\alpha
\\
&=&
\int_0^{\pi/2}
\left[
\frac{
\sin\left(
\frac{1}{2}QL\cos\alpha
\right)
}{
\frac{1}{2}QL\cos\alpha
}
\right]^2
\sin\alpha
\,d\alpha
\\
& & \text{substitute } u=\cos\alpha
\\
&=&
\int_0^1
\left[
\frac{
\sin\left(
\frac{1}{2}QLu
\right)
}{
\frac{1}{2}QLu
}
\right]^2
du
\\
& & \text{substitute } a=\frac{QL}{2}
\\
&=&
\int_0^1
\left[
\frac{
\sin(au)
}{
au
}
\right]^2
du
\\
& & \text{substitute } t=au
\\
&=&
\frac{1}{a}
\int_0^a
\left[
\frac{
\sin t
}{
t
}
\right]^2
dt
\\
& & \text{substitute }
\mathrm{Si}(x)=\int_0^x \frac{\sin t}{t}dt
\\
&=&
\frac{1}{a}
\left[
\mathrm{Si}(2a)
-
\frac{\sin^2 a}{a}
\right]
\\
& & \text{reinsert } a=\frac{QL}{2}
\\
&=&
\left[
\frac{
2\mathrm{Si}(QL)
}{
QL
}
-
\frac{
4\sin^2(QL/2)
}{
(QL)^2
}
\right]
\\
& & \text{substitute }
4\sin^2\left(
\frac{QL}{2}
\right) = 2 (1-\cos QL)\\
& = & \left[ \frac{2 Si(QL)}{QL} - \frac{2(1-\cos QL)}{(QL)^2} \right]
\end{eqnarray}

This is the form factor for a very thin cylinder of length $L$.

\subsection{Circular Disc}
\index{Form factor!Disc}

For the calculation of the disc form factor we again apply cylindrical coordinates. Following the convention that $f(Q)$ is the form factor amplitude and $F(Q)=|f(Q)|^2$ the corresponding form factor amplitude is [1]

\begin{eqnarray}
f(\vec{Q})
&=&
\frac{1}{V\Delta\rho_b}
\mathcal{F}(\rho_b(\vec{r}))
\\
&=&
\frac{1}{V\Delta\rho_b}
\rho_{b,\mathrm{contrast}}
\int_V
\exp\left(
-i\vec{Q}\cdot\vec{r}
\right)
\,dV
\\
& & \text{thin disc approximation } dV=r\,dr\,d\phi\cdot t
\\
&=&
\frac{t}{V}
\int_0^{2\pi}
\int_0^R
\exp\left(
-iQ_{\perp}r\cos\phi
\right)
r
\,dr\,d\phi .
\end{eqnarray}

where $Q_\perp = Q\sin\alpha$ and $\alpha$ is the angle between  $\vec{Q}$ and the disk normal, so that $\alpha = 90^\circ$ means we are looking directly at the disc face and therefore $\vec{Q}\vec{r}=Q_\perp r\cos \phi$. This means, $\phi$ is the angle that would rotate the disk from a flat to a side view.

We again separate directions, now radial and angular, since the disk is considered to be very thin with thickness $t$. We start with the angular integration over $\phi$ using a Bessel function

\begin{eqnarray}
\int_0^{2\pi}
\exp(-ix\cos\phi)
\,d\phi
&=&
2\pi J_0(x)
\\
& & \rightarrow
\\
f(Q,\alpha)
&=&
\frac{2\pi t}{V}
\int_0^R
rJ_0(Q_\perp r)
\,dr.
\end{eqnarray}

and the radial integration with the increase of the order of the Bessel function by integration

\begin{eqnarray}
\int
rJ_0(ar)
\,dr
&=&
\frac{
rJ_1(ar)
}{
a
}
\\
& & \rightarrow
\\
f(Q,\alpha)
&=&
\frac{2\pi t}{V}
\left[
\frac{
rJ_1(Q_\perp r)
}{
Q_\perp
}
\right]_0^R
\\
&=&
\frac{2\pi t}{V}
\frac{
RJ_1(Q_\perp R)
}{
Q_\perp
}
\\
&=&
2
\frac{
J_1(QR\sin\alpha)
}{
QR\sin\alpha
}.
\end{eqnarray}

This is the scattering function of a thin circular disc with $\alpha$ the angle between the surface normal of the disk and $\vec{Q}$. This is exactly what we would expect, since the fraction in the second term goes to 1 for $\sin \alpha = 0$ and therefore it would look like a round disk.

For creating the average over all disk orientations, we create the scattered intensity from the amplitude

\begin{eqnarray}
F(Q,\alpha)
&=&
|f(Q,\alpha)|^2
\\
&=&
4
\left[
\frac{
J_1(QR\sin\alpha)
}{
QR\sin\alpha
}
\right]^2 .
\end{eqnarray}

The orientational average then becomes

\begin{eqnarray}
F(Q)
&=&
\int_0^{\pi/2}
F(Q,\alpha)
\sin\alpha
\,d\alpha
\\
&=&
4
\int_0^{\pi/2}
\left[
\frac{
J_1(QR\sin\alpha)
}{
QR\sin\alpha
}
\right]^2
\sin\alpha
\,d\alpha
\\
&=&
\frac{2}{Q^2R^2}
\left(
1-
\frac{
J_1(2QR)
}{
QR
}
\right).
\end{eqnarray}

This is the orientation averaged form factor of a thin disc with radius $R$.

\subsection{Non-particulate scattering from a flexible chain}
\index{Form factor!Flexible Chain}
Since a flexible polymer chain does not have a geometric form that could be used for the Fourier transform here the approach is to derive the intensity directly from the pair correlation of scattering centers between particles $i$ and $j$ along the chain:

\begin{equation}
\left\langle (\vec{r}_i-\vec{r}_j)^2\right\rangle = |i - j|a_p^2
\end{equation}

with the persistence length $a_p$ between two segments.

\begin{eqnarray}
\left\langle \exp(-i \vec{Q}(\vec{r}_i-\vec{r}_j)) \right\rangle 		& = & \exp\left( -\frac{Q^2}{6}\langle(\vec{r}_i-\vec{r}_j)^2\rangle\right) \\
				&    &  \rightarrow \\
I (Q) 			& = & b^2 \sum_{i,j}\exp \left( -\frac{Q^2a_p^2}{6} |i-j|\right).
\end{eqnarray}

We are turning this sum into a double integral over $i$ and $j$ to $s$ and $s'$ variables with identical scattering length $b$ for all scattering centers

\begin{equation}
I(Q) = b^2 \int _0 ^N \int_0 ^N \exp (-a |s-s'|)ds ds'
\end{equation}

with $a=\frac{Q^2 a_p^2}{6}$. From that we exploit the symmetry

\begin{equation}
I(Q) = 2 b^2 \int _0 ^N ds \int_0 ^N \exp (-a |s-s'|) ds'
\end{equation}

and replace the distance between the two scattering centers by a constant $u = s-s'$

\begin{equation}
I(Q) = 2 b^2 \int _0 ^N ds \left[ \int_0 ^s \exp (-a u) du \right] ds.
\end{equation}

Evaluation of the integral in the bracket gets

\begin{equation}
\int _0 ^s \exp(-a u)du = \frac{1-\exp(-as)}{a}
\end{equation}

which leads to

\begin{eqnarray}
I(Q) 	& = & \frac{2b^2}{a}\int _0 ^N (1-\exp(-as))ds\\
	& = & \frac{2b^2}{a}\left(N- \frac{1-\exp(-aN)}{a}\right)\\
	& = & 2b^2 \left(\frac{N}{a}- \frac{1-\exp(-aN)}{a^2}\right)\\
	& = & 2b^2 N^2 \frac{\exp(-a N)-1+aN}{(aN)^2}\\
	&    &\text{substitute radius of gyration} R_g = \frac{N a_p^2}{6}\\
	&    &\text{and previous resubstitution} aN = Q^2 R_g^2\\
	& = &I(Q=0)\cdot \frac{2 (\exp(-Q^2 R_g^2)-1+Q^2R_g^2)}{(Q^2 R_g^2)^2}
\end{eqnarray}

which is the Debye scattering law:

\begin{equation}
I(Q)  = I(Q=0)\cdot \frac{2 (\exp(-Q^2 R_g^2)-1+Q^2R_g^2)}{(Q^2 R_g^2)^2} = \rho\Delta \rho^2 V^2 F_{\textrm{Debye}}(Q)
\label{eq:DebyeLaw}
\end{equation}

with $I(Q=0)=b^2N^2 = \Delta\rho_b^2 V^2$.

\subsection{Polydispersity}
All analytic form factors, that deliver the scattered intensity, are determining the scattered intensity for particles of one exact size. In real systems, however, there are mostly distributions of different sizes. This leads to a superposition of scattering from different particle sizes. Since most particle sizes follow a Gaussian distribution, this is also a good way to fold in the particle size distribution analytically. For extremely long, or very polydisperse, particles then Schulz-Zimm distribution is used, which looks very similar to the Gaussian distribution, however has a cut-off at zero to prevent negative sizes of the particles. For specialized problems also other distributions, such as La-Place, multi-modal or other size distribution functions can be used.

The general idea is that the scattered intensity $I(Q,r)$ is folded with the size distribution function $f(r)$

\begin{equation}
I_{\mbox{\small{real}}}(Q,r) = I_{\mbox{\small{ideal}}} (Q,r) * f(r).
\end{equation}

Here the subscripts real and ideal identify the real measured intensity or the ideal intensity for any calculated particulate size and form.

\begin{figure}
\centering
\includegraphics[width=0.75\textwidth]{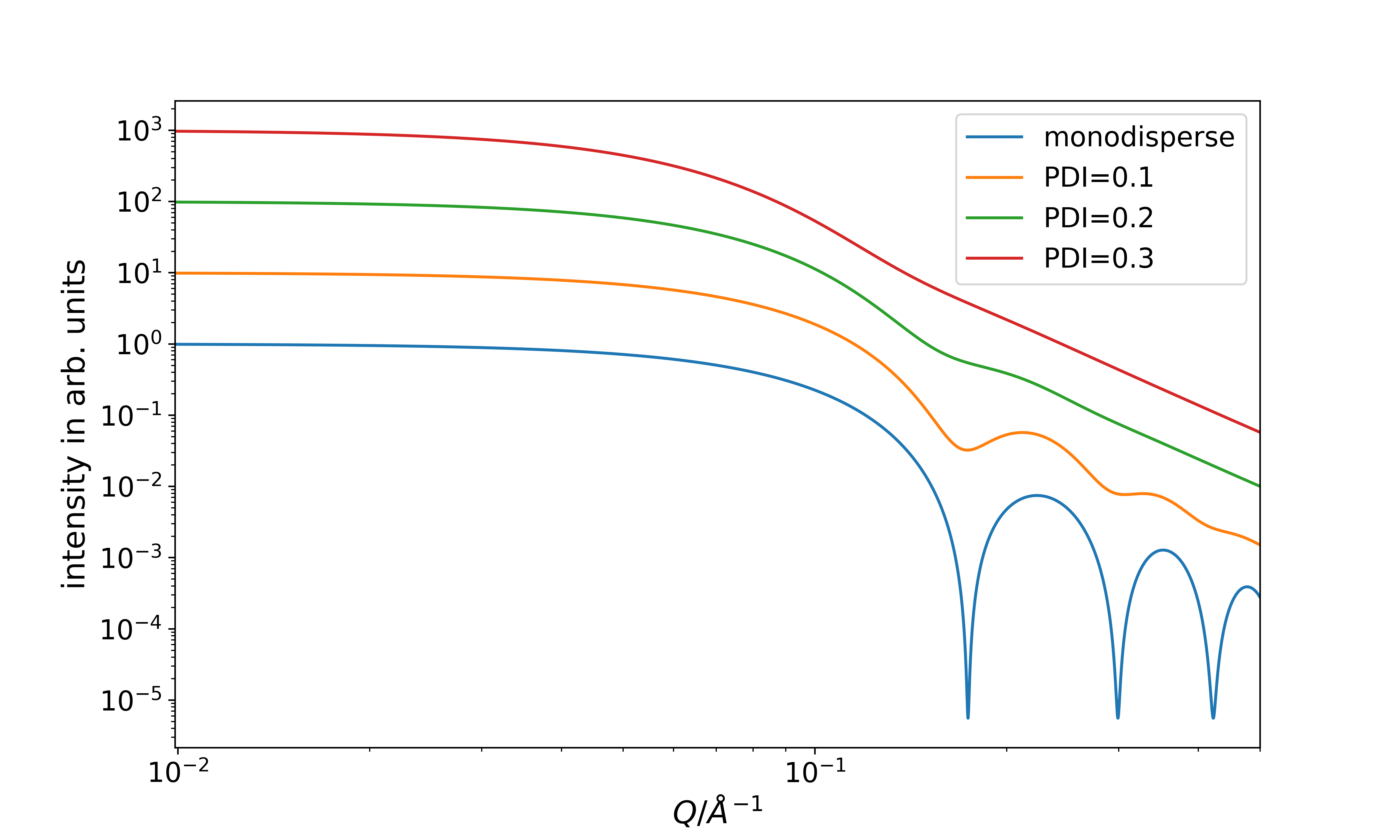}
\caption{Effect of polydispersity. While the positions of the minima can still be found at higher polydispersity, the higher order undulations of the form factor vanish.}
\label{fig:PDI}
\end{figure}

The effects of the convolution can be seen in Fig.\ref{fig:PDI}. Most notably, the minima are smeared out, and in some cases vanish completely, so they can only be estimated. Another important effect is that the slopes of inclinations cannot be completely reproduced anymore, which is especially important to distinguish scattering from different contributions. The magnitude of the polydispersity is described by the polydispersity index $PDI = \sigma(f(r))/\mu(f(r))$ where $\sigma(f(r))$ is the standard deviation of the size distribution function and $\mu(f(r))$ is the mean of the size distribution function. Values of $PDI \geq 0.3$ are usually discarded during fitting, as then the results become unreliable in such a polydisperse sample.

In addition to this, the usual polydispersity (approximated by a Gaussian distribution) is by its very nature similar to a resolution smearing of the instrument itself. Therefore, it can easily happen to overestimate the polydispersity. If the resolution function of the instrument is known, it should be used for deconvolution before performing the fits.

\section{Structure Factors}
Structure factors in general describe the scattered intensity due to the arrangement of single particles. This can be because the solution is becoming to dense, and therefore the particles arrange following a nearest neighbor alignment or because the particles are attractive to each other and form aggregates. Thus, more generally a structure factor $S(Q)$ is a measure of interaction between the single particles in the solution and connected with the correlation function $c(r)$ (the probability to find a particle at a certain distance) and the particle density $\rho$ with the relation

\begin{equation}
S(Q) = 1+
\frac{1}{N}
\sum_{i,j \,\mathrm{not\,in\,same\,particle}}
\exp(-i\mathbf{Q}\cdot\mathbf{r}_{ji})  = \frac{1}{1-nc(Q)}.
\label{eq:StructureFactor}
\end{equation}

\begin{figure}
\centering
\includegraphics[width=0.75\textwidth]{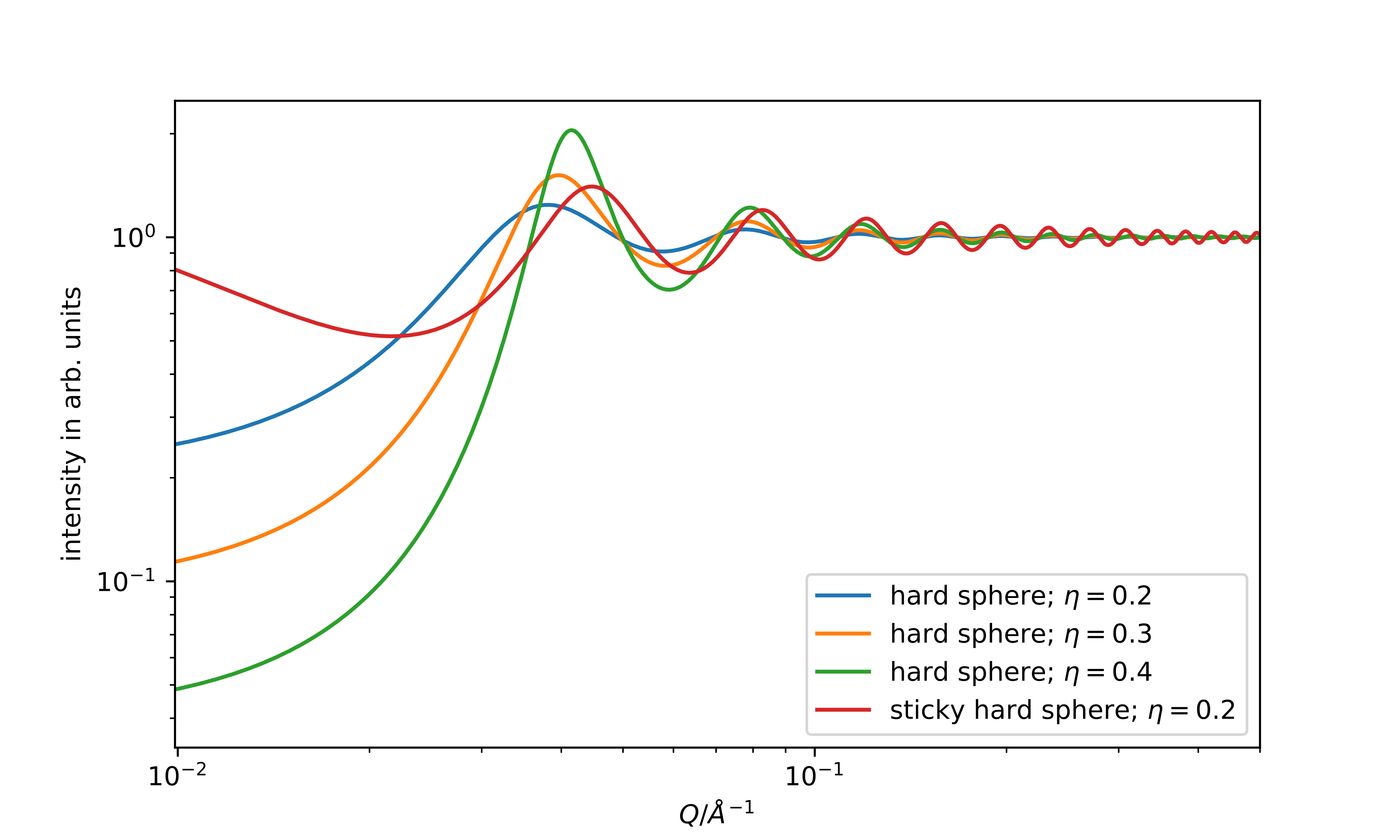}
\caption{Examples for structure factors. The intensity of the peaks roughly scales with the volume fraction $\eta$ of the particles. Also the position of the peaks is slightly dependent on that volume fraction, which makes a direct calculation of $R=\frac{2\pi}{Q_{max}}$ invalid (The hard sphere radius used here was 60 \AA). A distinct difference can be noted at low $Q$. Here, in general, attractive interaction (sticky hard spheres) leads to an increase in scattering, while repulsive interaction leads to a decrease in intensity.}
\label{fig:structureFactors}
\end{figure}

From this equation it also follows, that for a system of uncorrelated, identical particles the structure factor must be $S(Q)=1$, or in the case of a very low particle density. Since the correlation between particles usually leads to either an aggregation or repulsion of particles over long length scales the contribution of the structure factor is most prominent at low $Q$ values. Also, this means that for large distances the structure factor has to level out to unity, to preserve the fact that at large $Q$ only the inner structure of the particle is visible, not its arrangement in space. A few instructive examples for the structure factor are shown in Fig.\ref{fig:structureFactors}.

\subsection{Hard Sphere Structure Factor}
\index{Structure Factor}
The hard sphere structure factor assumes an infinitely high potential below a radius $R$ and a zero potential at higher radii. This can be described by

\begin{equation}
V(r) = \begin{cases}
	\infty \mathrm{\, for \,} r \leq R \\
	0 \mathrm{\hspace{3mm} for\, } r>R.
\end{cases}
\label{eq:HSPotential}
\end{equation}

Using Eq.\ref{eq:StructureFactor} this can be rewritten as

\begin{equation}
S(Q)=\frac{1}{1+24\eta_{HS}G(2QR)/2QR}.
\end{equation}

Here $G(x)$ is defined as

\begin{eqnarray}
G(x) & = & \alpha\frac{(\sin(x)-xcos(x))}{x^2}+\\
	& = & \beta\frac{(2x\sin(x)+(2-x^2)\cos(x-2))}{x^3}+\\
	& = & \gamma\frac{(-x^4\cos(x)+4\left[(3x^2-6)\cos(x)+(x^3-6x)\sin(x)+6\right])}{x^5}
\end{eqnarray}

with these definitions for $\alpha, \beta$ and $\gamma$:

\begin{eqnarray}
\alpha &=&\frac{(1+2\eta_{HS})^2}{(1-\eta_{HS})^4}\\
\beta&=&\frac{6\eta_{HS}(1+\small{\eta_{HS}}/\small{2})^2}{(1-\eta_{HS})^4}\\
\gamma&=&\frac{\small{\eta_{HS}}/\small{2}(1+2\eta_{HS})^2}{(1-\eta_{HS})^4}.
\end{eqnarray}

In all equations the volume fraction that is occupied by hard spheres of radius $R$ is designated $\eta_{HS}$.

\section{Reading a curve}
\begin{figure}
\centering
\includegraphics[width=0.75\textwidth]{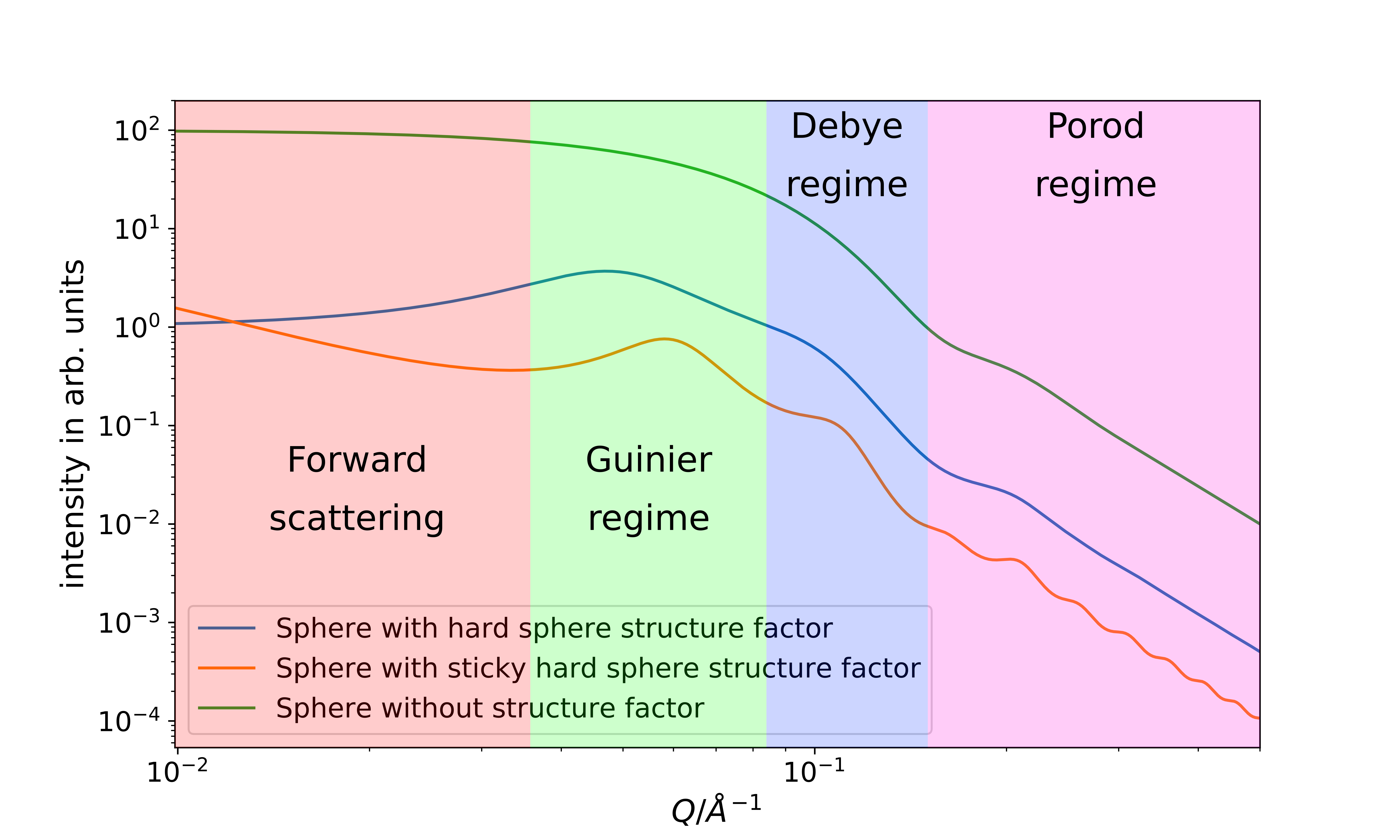}
\caption{Diverse scattering curves from identical spherical form factor and different structure factors.}
\label{fig:curveReading}
\end{figure}

In an experimental environment it can be useful to determine the fundamental features in a preliminary fashion without computer aided data evaluation, also known as fitting. In addition, this helps determining good starting parameters for fits. In order to do so, we are going to look at the curves shown in Fig.\ref{fig:curveReading}. There we can determine different regions of the scattered intensity (forward scattering, Guinier regime, Debye regime and Porod regime) and determine several properties of the sample from that intensity. When applying the described techniques for directly reading a curve it has to be kept in mind that most of them are either restricted in their validity to specific regions of $Q$ space or are very general and rough descriptions of the sample.

\subsection{Forward scattering}
As pointed out in the discussion of the structure factor, large aggregates are typically characterized by an increased scattering intensity at low $Q$. This behaviour is also evident from Eq.\ref{eq:distances} into account. This means, in general, an increased scattering at low $Q$ is indicative of large aggregates being present in the sample. This also correlates with an attractive potential between the single particles.

Another possibility is strongly suppressed scattering at low $Q$. This can be the case for strongly repulsive interaction potentials between the particles, close to what is described for the hard sphere factor above.

A levelling out of the intensity at low $Q$ is indicative of an either dilute solution or a very weak potential between the particles. Then there is no influence at low $Q$ and only the structure factor of the single particles is visible.

\subsection{Guinier regime}
\label{sec:GuinierRegime}
\index{Guinier Law}
The Guinier regime is usually the crossover region, where the forward scattering is not dominant anymore and the slope of the scattering curve changes to the scattered intensity of the form factor. In this regime the overall size of the particle can be examined. This is similar to seeing something from far away: One may be able to discern the size of the particle but the distinct form remains hidden. Imagine a football and a pumpkin seen from 100 m away. They are close in size, you can properly judge it to be approximately 20 cm in diameter, but the exact form (ridges, stem of the pumpkin) remains hidden. A description that is only taking into account the scattered density of the particles as a whole, valid in that scattering regime is the Guinier Law:

\begin{equation}
I(Q)=\Delta\rho_b V^2\exp(-\frac{Q^2R_g^2}{3})
\label{eq:Guinier}
\end{equation} 

For details of derivation, which include a Taylor series expansion around zero of the scattered amplitude (Eq.\ref{eq:GeneralFT}) and an averaging over all directions, please refer to the literature.\cite{guinier,roe} Another option is to develop a series expansion for the Debye Law (Eq.\ref{eq:DebyeLaw}) at low $Q$.

\begin{figure}
\centering
\includegraphics[width=0.75\textwidth]{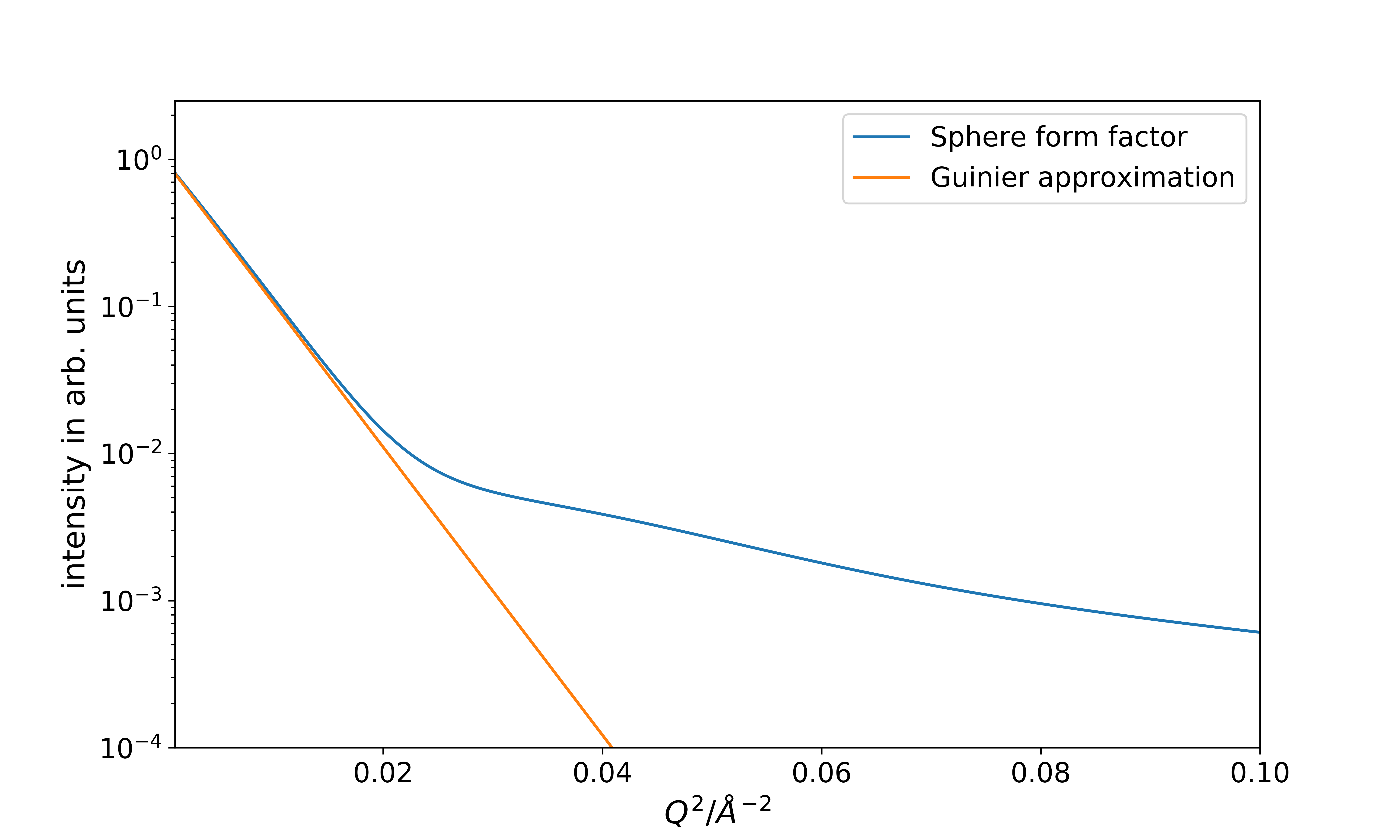}
\caption{Sphere form factor and Guinier approximation from Eq.\ref{eq:Guinier} in a Guinier plot. The radius of gyration is 25.8 \AA. The estimated slope by eye was $m=-185$. With $R_g=\sqrt{-3\cdot m} = 23.5$\AA\, the error is within 10\%, which is suitable for a naked eye approximation.}
\label{fig:GuinierPlot}
\end{figure}

In order to evaluate the data using the Guinier Law, the data needs to be plotted as shown in Fig.\ref{fig:GuinierPlot}. The log-log representation and plotting versus $Q^2$ allow to directly read the inclination of the system, multiply by 3 and use the square root in order to retrieve the particle radius.

\subsection{Debye regime}
In contrast to the Guinier regime, where the data can be evaluated by the Guinier law, the Debye regime signifies the area, where the particulate form manifests in the scattering, which in general cannot be fitted by the Debye law. The Debye law is only valid for the scattering from Gaussian chains. As can be seen in the form factors section \ref{sec:formFactors}, there is a direct correlation to the dimensionality of the scattering particle (sphere, disc, rod) and the slope in log-log plot, since the scattering scales with $I(Q)\sim Q^{-D}$, where $D$ is the dimensionality of the scattering object (sphere: $D=3$; disc: $D=2$; rod: $D=1$). Also the scattering from fractal objects is possible, which then results in non-integer numbers for the slope. It should be noted that this is an approximation that is only valid for the case when $1/\mbox{particle radius} \ll Q \ll 1/\mbox{fundamental building block}$. The fundamental building block in this case can be for example atoms or single monomers of a chain.

\subsection{Porod regime}
\index{Porod Law}
The Porod regime, is the regime where the interface between the particle and the solvent dominates the scattered intensity. It is valid for large $Q$ (before leveling out into the incoherent background) and therefore a good approach is extrapolating the sphere form factor to large $Q$. The decisive property of the scattered intensity is the scaling of $I(Q)\sim Q^{-4}$. This behavior can be derived from an extrapolation of the sphere form factor (Eq.\ref{eq:SphereForm}) to very large Q:

\begin{eqnarray}
I(Q) &\propto & \left( \frac{4}{3}\pi R^3\right)^2\frac{9 (\sin QR -QR\cos QR)^2}{Q^6 R^6}\\
	& = & 8\pi ^2\left( \frac{R^2(1+\cos 2QR)}{Q^4}-\frac{2R\sin 2QR}{Q^5}+\frac{1-\cos 2QR}{Q^6}\right)
\label{eq:SpherePorod}
\end{eqnarray}

The higher order terms vanish at large $Q$ delivering the characteristic $Q^{-4}$ behavior of the scattered intensity. Here only proportionality is claimed, which is strictly true in this case. If the scattered intensity is recorded in absolute intensities, here also information about the surface of the particles can be obtained. This then follows the form

\begin{equation}
\lim _{Q\rightarrow \infty}I(Q)=\frac{8\pi \Delta\rho_b S}{Q^4}.
\label{eq:Porod}
\end{equation}

$ \Delta\rho_b$ is here the SLD difference between the particle and the surrounding medium and $S$ the inter-facial area of the complete sample between particles and medium. This means, the absolute intensity of the Porod regime allows to determine the complete amount of surface in the sample.

\subsection{Estimation of particle and feature Size}
As described previously for low $Q$ in most cases it is a good approximation to assume all particles in the sample have spherical symmetry (Section \ref{sec:GuinierRegime}). The roots of the expression for the spherical form factor are in the locations $\tan (QR)=QR$, which is true for $QR\approx 4.49, 7.73, 10.90 ....$. In many cases anyway only the first minimum of the form factor will be visible. This allows a fast approximation of the radius with $R\approx 4.5/Q_{\text{min}}$. Here it needs to be noted, that this is the rotational average of the particle, neglecting any structure of the particle whatsoever.

Another approach of determining the size or correlation of features is using Eq.\ref{eq:distances}:

\begin{equation}
d = \frac{2\pi}{Q}.
\end{equation}

Although this is in general only strictly true for lamellar systems and the corresponding correlations, it is still a good approximation for a summary data examination during the experiment. With that restriction in mind it can be used for virtually any feature in the scattering curve and the size of the corresponding feature in the sample.

\section{Further Reading}
Most of the concepts shown in this manuscript are based on previous publications. The following selection of textbooks gives the reader a good overview of the principles of SAS.

\subsection{A. Guinier: X-ray diffraction in crystals, imperfect crystals, and amorphous bodies}
This early textbook concentrates on SAXS, as neutron scattering at the time of writing was still in its infancy. While some of the terminology may have changed slightly over time, in many aspects this book still gives a good fundamental overview of what can be done with small-angle scattering, and how to perform a solid data analysis. In addition, this is literally the book on the Guinier Law, and where some of the basic ideas of reading scattering curves were first collected.

\subsection{R.J. Roe: Methods of x-ray and neutron scattering in polymer science}
Here the author nicely manages to emphasize the commonalities and differences between x-ray and neutron scattering. An overview of the methods and technologies is given, as well as a helpful mathematical appendix, reiterating some of the concepts used in the book.

\subsection{G. Strobl: The physics of polymers}
For soft-matter researchers this book, even though not being focused on scattering as such, gives a good overview of applicable concepts for scattering with soft-matter samples. A wide range of helpful examples highlight in which particular area any evaluation concept of the data is applicable and useful.

\section[APPENDIX: SCATTERING LENGTH CALCULATION]{Appendix: Calculation of the Scattering Length by solution of Schrödinger's equation}
\label{SchroedingerAppendix}
\fancyhead[LE]{Small-Angle Neutron Scattering}
\fancyhead[RO]{\nouppercase{\leftmark}}

In order to calculate the scattering length density, we have to start with the time-independent Schrödinger equation

\begin{equation}
-\frac{\hbar ^2}{2 m}\nabla ^2\psi + V \psi = E \psi.
\end{equation}

Using separation of variables as shown in quantum mechanics textbooks \cite{IntroQM} we can create a radially symmetric version of Schrödinger's equation with 

\begin{equation}
\psi(r, \theta, \phi) = R(r)Y(\theta, \phi)
\end{equation}

where the radial part of the wave function is captured by $R(r)$. This allows us to rewrite Schrödinger's equation as 

\begin{equation}
-\frac{\hbar ^2}{2 m}\frac{d^2u}{dr^2}+\left[V+ \frac{\hbar ^2}{2 m}\frac{l(l+1)}{r^2} \right] u = E u.
\label{eq:SchrodingerRadial}
\end{equation}

Here we substituted $u(r) = r R(r)$, $R = u/r$, $dR / dr = \left[ r(du/dr) - u\right]/r^2$ and $(d/dr)\left[ r^2()dR/dr\right] = rd^2u/dr^2$. $l$ is the quantum number for angular momentum, and this is called the radial Schrödinger equation. $m$ in this instance are still masses, not quantum numbers. 

For this we now only account for solutions $V(r)$, since we assume the nuclei to be radially symmetrical. For non-symmetric solutions $V(r,\theta, \phi)$ this becomes more complicated and in the case of neutrons is exclusively interesting for magnetic scattering or high energy interaction. Magnetic scattering is using another potential, which is often weak combined with the nuclear interaction. Higher energy interactions are not the focus of this course, since for neutron scattering the energies for thermal (25 meV) and cold (1 meV) neutrons always are too low to reach higher excitations the $l=0$ and s-wave scattering. This is a sensible assumption since we will only be scattering with cold neutrons at low energies. This also implies that the potential is a lot deeper than the energy of the incoming neutron $E \ll V_0 $. So let us consider a potential of the form

\begin{equation}
V(\vec{r}) =
\left\{
\begin{array}{ll}
-V_0 & \text{if } r \leq \text{radius } a\\
0 & \text{if } r > \text{radius } a
\end{array}
\right.
\end{equation}

All this leads to neutrons interacting with the nuclei directly, which results in the atomic form factor being always spherically symmetric (billiard balls) and them being sensitive to different isotopes and spin-spin coupling. In contrast to x-rays, there is no simple expression for scattering strength as a function of isotope or atomic number. Directly neighbouring elements and isotopes may have vastly different cross sections.

This is due to the fact that the Schr\"odinger equation has to be solved for each combination of incoming neutron and nucleus. The solution for the problem is illustrated by Hammouda or Tong \cite{hammouda1995tutorial, ScatteringTheory} in more detail. 

In order to achieve a solution, we consider the continuity for the wave-function and its derivative at the boundaries. The radial Schrödinger equation for $l=0$ for potential $V_0$ inside the well is

\begin{equation}
\frac{d^2u}{dr^2} + \left[ k_1^2 -\frac{1}{r^2}\right] u = 0
\end{equation}

With $k_1^2 =\frac{2m(E+V_0)}{\hbar ^2} $. The general solution for this is

\begin{equation}
u_{I} = A j_l (k_1 r) + B n_l (k_1 r)
\end{equation}

with $j_l$ being the sperical Bessel function and $n_l$ being the spherical Neumann function. For $r=0$ the Neumann function goes to infinity, which is unphysical, so $B=0$. So the general solution inside the well is

\begin{equation}
u_{I} = A j_l (k_1 r).
\end{equation}

Outside the well $V=0$ so $k^2 = \frac{2mE}{\hbar ^2}$ which renders the general solution

\begin{equation}
u_{II} = C j_l (k r) + D n_l (k r).
\end{equation}

Since $r\neq 0 $ in all cases, all parameters are allowed.

We consider positions far from the origin, where the particle is free

\begin{equation}
j_{0} (kr) \sim \frac{1}{kr}\sin \left( kr \right)
\end{equation}

and 

\begin{equation}
n_{0} (kr) \sim \frac{1}{kr}\cos \left( kr \right)
\end{equation}

which leads to

\begin{equation}
u_{II} \sim \sin \left( kr +\delta_0\right)
\end{equation}

The phase shift encodes the effect of the potential well as different from a completely free particle with

\begin{equation}
\tan \delta_0 =  -\frac{D}{C}.
\end{equation}

With the continuity condition at $r = a$ we get

\begin{equation}
u_{I} (a) = u_{II} (a)
\end{equation}

meaning

\begin{equation}
A j_0 (k_1 a) = C j_0 (ka) + D n_0 (ka).
\end{equation}

In addition the derivative needs to be continuous

\begin{equation}
A j_0' (k_1 a) = C j_0' (ka) + D n_0' (ka).
\end{equation}

The $l=0$ condition gives

\begin{equation}
j_0 (x) = \frac{\sin x}{x}
\end{equation}

and 

\begin{equation}
n_0 (x) = -\frac{\cos x}{x}
\end{equation}

which results in the general solutions

Inside:

\begin{equation}
u_I (r) = A \sin (k_1 r)
\end{equation}

and outside:

\begin{equation}
u_{II} (r) = C \sin (kr) + D \cos (kr).
\end{equation}

We remember the small perturbance trick with the $\delta$ from the general solution and get

\begin{equation}
u_{II} (r) = F \sin (kr + \delta_0) = F \cos \delta_0 \sin (kr)+F \sin \delta _0 \cos(kr)
\end{equation}

meaning $C = F \cos \delta_0$ and $D = F \sin \delta _0$ which is equivalent to our previous definition with $\tan \delta _0 = D/C$. The difference in sign is due to the asymptotic nature of the Neumann function, which by convention is then written negative. For the sine and cosine solution used here, the positive $\delta _0$ as a small perturbation is the convention.

Again, applying the continuity condition for both the solution itself and its derivative at $r=a$ we get

\begin{equation}
A\sin (k_1 a) = F\sin(k a+\delta _0)
\end{equation}

for the solution itself and

\begin{equation}
A k_1 \cos(k_1 a) = F k \cos (k a+ \delta _0)
\end{equation}

for its derivative.

Now we have two functions containing $\delta _0$ we can use to solve for it. Dividing the derivative of the function by the function itself gives

\begin{equation}
\frac{A k_1 \cos(k_1 a)}{A\sin (k_1 a)}=\frac{F k \cos (k a+ \delta _0)}{F\sin(k a+\delta _0)} = k_1 \cot (k_1 a) = k \cot (ka \delta _0).
\end{equation}

We solve this for the phase shift $\delta _0$. We remember the complete perturbation (the nuclear core, which is responsible for the actual scattering) is contained in there.

\begin{equation}
\cot (ka +\delta _0) = \frac{k_1}{k}\cot (k_1 a)
\end{equation}

which leads to

\begin{equation}
\delta _0 = - ka +\tan ^{-1} \left( \frac{k}{k_1}\tan (k_1 a)\right).
\end{equation}

In the literature you will find this usually as the s-wave phase shift for a finite spherical square well.

In order to interpret this physically we remember the definitions of the different $k$'s being $k = k_1 (V_0 = 0)$. In physical terms this means that the solution inside and outside are only different in the way, that outside the well ($k$-solution) the potential is zero, whereas inside the well ($k_1$-solution) the potential is finite negative $-V_0$.

Looking at the solution for delta, if there was no potential at all, $k_1 = k$ would be true in all cases, so would be $k_1/k =1$. This leads to 

\begin{equation}
\delta _0 = -ka + \tan^{-1}\left(\tan(ka)\right) = -ka + ka = 0
\end{equation}

which is exactly what we expect. For very low energies $ka \ll 1$ we can approximate $\delta _0 \approx -ka_s$, with $a_s$ chosen as a radius for low energy, s-wave scattering:

\begin{equation}
\delta _0 = -k a_s = - k a + \frac{k}{k_1}\tan(k_1 a)
\end{equation}

which directly leads to an expression for the scattering length $a_s = b$, which is exactly the radius of a given weak potential:

\begin{equation}
a_s = b = a - \frac{\tan (k_1 a)}{k_1}.
\end{equation}

\newpage
\addcontentsline{toc}{section}{References}
\bibliographystyle{ieeetr}
\bibliography{example_nlab}

\end{document}